\documentclass[11pt,a4paper]{article}

\usepackage{amsbsy}
\usepackage{amssymb}
\usepackage{graphicx}
\usepackage[]{natbib}
\usepackage{booktabs}
\usepackage{longtable}
\usepackage{threeparttable}
\usepackage{bigstrut}
\usepackage{setspace}
\usepackage{placeins}
\usepackage{pdflscape}
\usepackage{eurosym}
\usepackage{amsmath}
\usepackage{mathtools}
\usepackage[utf8]{inputenc}
\usepackage{lmodern}
\usepackage{enumerate}
\usepackage{float}             
\usepackage{makecell}
\usepackage{relsize}
\usepackage{color} 
\usepackage{listings}
\usepackage{hyperref}
\usepackage{comment}
\definecolor{Mygrey}{gray}{0.75}
\definecolor{Cgrey}{gray}{0.4}
\newtheorem{definition}{Definition}[section]

\newtheorem{theorem}[definition]{Theorem}
\newtheorem{prop}[definition]{Proposition}

\numberwithin{equation}{subsection}

\graphicspath{{Pictures/}}

\usepackage[left=2.54cm, right=2.54cm, top=2.54cm, bottom=2.54cm]{geometry}

\newcommand{\blind}{0}

\begin{document}

\def\spacingset#1{\renewcommand{\baselinestretch}%
	{#1}\small\normalsize} \spacingset{1}
\spacingset{1.84}
	
\bibliographystyle{chicago}


\if0\blind
{
	\title{\textbf{Sequentially valid inference for probabilistic inflation forecasts}}
\author{
\renewcommand{\thefootnote}{\alph{footnote}}
\textsc{Amadeo Grob} \footnotemark[1]
\and
\renewcommand{\thefootnote}{\alph{footnote}}
\textsc{Maurizio Daniele} \footnotemark[2]
\and 
\renewcommand{\thefootnote}{\alph{footnote}}
\textsc{Johanna Ziegel} \footnotemark[3]
}
\date{\today}
\renewcommand{\thefootnote}{\alph{footnote}}
\footnotetext[1]{Corresponding Author. University of St. Gallen, Mathematics and Statistics Division, Rosenbergstrasse 22, 9000 St. Gallen, Switzerland. Email: \href{mailto:amadeoelia.grob@unisg.ch}{amadeoelia.grob@unisg.ch}}
\footnotetext[2]{ETH Zürich, KOF Swiss Economic Institute, 8092 Zurich, Switzerland. Email: \href{mailto:daniele@kof.ethz.ch}{daniele@kof.ethz.ch}}
\footnotetext[3]{ETH Zürich, Seminar for Statistics, Rämistrasse 101, 8092 Zürich, Switzerland. Email: \href{mailto:ziegel@stat.math.ethz.ch}{ziegel@stat.math.ethz.ch}}
\maketitle
} \fi

\if1\blind
{
\bigskip
\bigskip
\bigskip
$ $
\vspace{1cm}
$ $
\begin{center}
	{\LARGE\bf Sequentially valid inference for probabilistic inflation forecasts} \\
	\bigskip
	\today
\end{center}
\medskip
} \fi

\bigskip
\begin{abstract}\footnotesize
Traditional statistical tests are poorly suited for the sequential evaluation of probabilistic forecast calibration. We address this limitation in macroeconomic forecasting by applying a new sequential testing method based on e-values. The e-value-based methodology enables anytime-valid inference. It allows practitioners to test against calibration continuously without invalidating statistical guarantees. To illustrate the framework's practical value, we apply it to probabilistic inflation forecasts for the United States, the Euro Area, and Switzerland. Our analysis shows that the sequential approach gives detailed insights into the timing and nature of forecast misspecification. We find these diagnostics are particularly insightful during major structural breaks. During these events, we find evidence against calibration that static, full-sample tests often miss. Therefore, this work shows that e-value-based tests are a practical method for the evaluation of forecast calibration in empirical macroeconomics.\\ 

\end{abstract}

\section{Introduction}\label{sec:intro}
Central banks rely on accurate inflation forecasts to guide monetary policy and communication. While point forecasts remain the most visible form of forecast communication, probabilistic forecasts provide a richer representation of uncertainty by describing the full distribution of possible future outcomes. In practice, however, major central banks differ in how they communicate this uncertainty. The Bank of England, for example, pioneered density forecasts by publishing inflation fan charts since 1997 \citep{bernanke_forecasting_2024, monetary_policy_committee_boe_monetary_2025}. In contrast, other central banks such as the US Federal Reserve (Fed), the European Central Bank (ECB), and the Swiss National Bank (SNB) provide less explicit distributional information. This variation in communication practices also reflects a broader gap in the evaluation of probabilistic inflation forecasts. Existing research has focused on the calibration of the Bank of England's fan charts, with some studies finding evidence of systematic miscalibration \citep{clements_evaluating_2004, wallis_assessment_2004}, while others provide a more favorable assessment \citep{elder_assessing_2005}. In contrast, relatively little empirical evidence exists on the calibration of density forecasts for other major central banks, including the Fed, ECB, and SNB \citep{fomc_minutes_2007, reifschneider_gauging_2007, chahad_update_2024, european_central_bank_eurosystem_2025}. This paper contributes to closing this gap by developing, monitoring, and evaluating probabilistic inflation forecasts for the United States, the Euro Area, and Switzerland. In particular, we focus on forecast calibration, a fundamental property of probabilistic forecasts that determines whether reported uncertainty accurately reflects the likelihood of future outcomes.

A probabilistic forecast is calibrated if observed outcomes are consistent with the forecast distributions. In other words, the forecast probabilities should coincide with the empirical frequencies of the realized events. A widely used diagnostic tool to assess calibration is the probability integral transform (PIT) \citep{diebold_comparing_1995, gneiting_probabilistic_2007}. The PIT is defined as the value of the forecast cumulative distribution function evaluated at the observed outcome. Under probabilistic calibration, the resulting PIT values are expected to be distributed according to a standard uniform distribution. In practice, calibration is commonly assessed using graphical diagnostics such as PIT histograms. A histogram that is approximately uniform provides evidence of well-calibrated forecasts, whereas systematic deviations from uniformity indicate calibration deficiencies. The shape of the histogram can also provide insight into the nature of the miscalibration. For example, a skewed histogram suggests systematic biases, while an inverse U-shaped histogram is indicative of overdispersion in the predictive distributions, meaning that the predictive distributions are excessively diffuse and therefore overestimate the true uncertainty.

Despite their widespread use, traditional calibration tools have a fundamental limitation: they are designed for fixed-sample settings and therefore ignore the sequential nature of forecast evaluation. In practice, particularly in macroeconomic forecasting, the model performance is monitored continuously rather than assessed only after a predetermined evaluation period has ended \citep{arnold_sequentially_2023}. Classical goodness-of-fit tests for PIT uniformity, such as the Kolmogorov–Smirnov test, are valid only when the evaluation sample size is specified \textit{ex ante}. This assumption is rarely satisfied in practice. Instead, analysts often evaluate the forecast performance whenever concerns about the model adequacy arise. Such a data-dependent monitoring, known as optional stopping, invalidates the nominal p-values of conventional hypothesis tests \citep{shafer_game-theoretic_2019, arnold_sequentially_2023}. Consequently, repeated testing over time inflates the probability of false rejections and undermines the statistical validity of classical calibration tests. A further limitation is that forecast calibration is rarely constant over time. Structural changes in the data-generating process may cause a forecasting model to alternate between periods of good calibration and periods of systematic miscalibration. Moreover, biases in opposite directions can even cancel out over an extended sample. For example, a period of under-prediction followed by over-prediction could yield a PIT distribution that looks deceptively uniform \citep{arnold_sequentially_2023}. A static, full-sample test may fail to detect important temporal changes in the forecast performance. These issues highlight the need for a new testing framework. This framework should remain remain valid under sequential testing, adapt to structural changes, and allow for real-time calibration assessment without violating statistical error rates.

To address these limitations, we adopt the sequential inference framework for forecast calibration proposed by \citet{arnold_sequentially_2023}. Rather than relying on classical p-values, this framework quantifies statistical evidence using e-values. An e-value is a non-negative random variable with expectation at most one under the null hypothesis \citep{shafer_testing_2021, arnold_sequentially_2023}. Consequently, large e-values provide evidence against the null. A key advantage of e-values is that they can be combined multiplicatively over time to form a test supermartingale under the null hypothesis, which guarantees anytime-valid inference \citep{grunwald_safe_2024}. Therefore, practitioners can use the test supermartingale to assess calibration at any point in time. This allows us to monitor forecast calibration continuously with valid Type~I error control, even with optional stopping. When the forecasts are well-calibrated, the test supermartingale is unlikely to attain large values. In contrast, persistent miscalibration causes the test supermartingale to grow, thereby accumulating evidence against the null hypothesis. This approach is therefore particularly well-suited for the real-time, adaptive environments, such as those encountered by central banks and other institutions that routinely update and evaluate predictive models. 

We construct a suite of probabilistic inflation forecasting models and evaluate the calibration of their predictive distributions using the sequential e-value framework. Our empirical analysis focuses on monthly headline inflation forecasts at horizons of 1, 3, 6, and 12 months for the United States, the Euro Area, and Switzerland. For each region, we generate predictive distributions using a diverse set of forecasting models, ranging from simple benchmark methods, such as naive forecasts and autoregressive (AR) models, to more sophisticated approaches, including Bayesian vector autoregressions (BVARs), a dynamic factor model (DFM), and the distributional random forest, a modern nonparametric forecasting method \citep{cevid_distributional_2022}. Forecast calibration is assessed using sequential e-value tests. These tests use PIT and rank values to check for continuous and discrete uniformity. To complement the calibration analysis, we also evaluate point and distributional forecast accuracy using conventional performance measures, including the root mean squared error (RMSE), mean absolute error (MAE), and the continuous ranked probability score (CRPS). Together, these metrics provide a comprehensive assessment of both forecast accuracy and probabilistic calibration. The sequential e-value framework enables continuous monitoring of the forecast performance over time. Our empirical results show that calibration deficiencies are concentrated in distinct episodes rather than persisting uniformly throughout the evaluation sample. Simple benchmark models, such as the Gaussian no-change forecast and univariate AR models, generally exhibit satisfactory calibration. In contrast, the BVAR models tend to be systematically overconfident, displaying pronounced calibration failures during periods of structural change, including the global financial crisis in the United States and the post-pandemic inflation surge in the Euro Area. Although the DFM performs more favorably than the BVAR specifications, the sequential diagnostics nevertheless reveal horizon-dependent calibration deficiencies that are not consistently detected by conventional full-sample tests. These findings illustrate the practical value of the e-value framework for real-time forecast evaluation. Beyond identifying whether a forecasting model is miscalibrated, the sequential approach pinpoints when calibration deteriorates and whether the deterioration is transient or persistent. This ability to detect and monitor time-varying changes in forecast calibration constitutes a key advantage over traditional fixed-sample evaluation methods.

The rest of the paper is organized as follows. Section~\ref{sec:evaluation_theory} introduces the theoretical framework for evaluating probabilistic forecasts, defines calibration in the sequential setting, and presents the e-value tests used in our analysis. Section~\ref{sec:forecast_design} describes the macroeconomic datasets, forecast design, and the considered forecasting models. Section~\ref{sec:results} reports our main empirical findings on forecast calibration across all regions, models, and horizons. Finally, Section~\ref{sec:conclusion} summarizes the results and discusses their implications.

\section{Evaluating probabilistic forecasts}
\label{sec:evaluation_theory}

First, we review classical methods for forecast evaluation, and then introduce e-values as a modern tool to test the calibration of probabilistic forecasts. Unlike classical tests, e-values remain valid if a forecaster monitors calibration continuously or stops testing at any point. The methods and definitions we present are based mainly on the work of \citet{arnold_sequentially_2023}.

\subsection{Notions of calibration}
\label{sec:notionscalibration}

We model the year-over-year inflation rate as a stochastic process $(Y_t)_{t \in \mathbb{N}}$ on a probability space $(\Omega, \mathcal{F}, \mathcal{P})$. The available information over time is represented by a filtration $(\mathcal{F}_t)_{t \in \mathbb{N}}$, a sequence of nested $\sigma$-algebras where $\mathcal{F}_t \subset \mathcal{F}_{t+1} \subset \mathcal{F}$. We assume the process $Y_t$ is adapted to the filtration, that is, $Y_t$ is $\mathcal{F}_t$-measurable for each $t$, or in other words, at time point $t$, we know the value of $Y_t$. A probabilistic forecast for $Y_{t+h}$ provides a full predictive probability distribution based on the available information $\mathcal{F}_t$ at the time of forecasting \citep{gneiting_probabilistic_2014}. This approach allows to explicitly quantify uncertainty of the future outcome.

Let $h \in \mathbb{N}$ be the forecast horizon. The true conditional cumulative distribution function (CDF) of a future value $Y_{t+h}$, given the information $\mathcal{F}_t$, is
\[
F_{t,h}(y) = \mathbb{P}(Y_{t+h} \le y \mid \mathcal{F}_t), \quad y \in \mathbb{R}.
\]
This distribution is the theoretical, unobservable ground truth. A probabilistic forecast, which we denote by the predictive CDF $\hat{F}_{t,h}$, is an estimate of this true conditional CDF. 
This predictive CDF, $\hat{F}_{t,h}$, is our central object of interest. In our application, we denote the realization of the random variable $Y_{t+h}$ by the standard econometric convention, $\pi_{t+h}$. We can derive any desired statistical functional from the predictive CDF. For example, a point forecast for the median is the 50\%-quantile, and a 90\% central prediction interval is the range $[\hat{F}_{t,h}^{-1}(0.05), \hat{F}_{t,h}^{-1}(0.95)]$, where $\hat{F}^{-1}_{t,h}$ denotes the generalized inverse, or quantile function. 

The quality of a probabilistic forecast is determined by two key properties: calibration and sharpness \citep{gneiting_probabilistic_2007}. Calibration refers to the statistical consistency between the predictive distributions and the observed outcomes. It is a joint property of the forecasts and the realizations. A calibrated forecast is ``statistically honest'' because its stated probabilities match the long-run frequencies of events. Sharpness, in contrast, is a property of the forecast alone. It refers to the concentration of the predictive distribution; narrower forecasts are sharper. Sharpness is essential for a forecast to be informative. The forecaster's goal is to achieve the greatest possible sharpness subject to maintaining calibration.

In this article, we focus on forecast calibration, more specifically on probabilistic calibration, which is arguably the most common notion of calibration considered in practice \citep{Dawid1984, diebold_evaluating_1998, gneiting_probabilistic_2007}. For a given predictive CDF $\hat{F}$, and a real-valued outcome $Y$, the probability integral transform (PIT) is defined as 
\[
Z_{\hat{F}}(Y) = \hat{F}(Y-) + V(\hat{F}(Y) - \hat{F}(Y-)),
\]
where $\hat{F}(y-) = \lim_{z \downarrow y} \hat{F}(z)$ is the left-limit of the CDF, and $V$ is an auxiliary random variable with $V \sim \text{UNIF}(0,1)$, independent of $(\hat{F}, Y)$. The forecast $\hat{F}$ is \emph{probabilistically calibrated} if its PIT is uniformly distributed on the unit interval, i.e., $Z_{\hat{F}}(Y) \sim \text{UNIF}(0,1)$. The randomization of the PIT through $V$ is necessary to handle potential discontinuities in the predictive CDF, ensuring that the resulting PIT is continuous. 


An ensemble forecast $\mathcal{X} = (X_1, \dots, X_m)$ is a collection of $m$ possible values of the future outcome $Y$, which are typically interpreted as random draws of the conditional distribution of $Y$. The calibration of ensemble forecasts is usually assessed through rank histograms, which are closely related to PIT histograms. For simplicity, we assume that the conditional distribution of $Y$ is continuous, which is a natural assumption in our application. 
Then, the rank of the outcome $Y$ with respect to the ensemble forecast $\mathcal{X}$ is
\[
\text{rank}_{\mathcal{X}}(Y) = 1 + \sum_{i=1}^m \mathbf{1}_{\{X_i < Y\}}.
\]
An ensemble forecast $\mathcal{X}$ is \emph{rank calibrated} if its rank is uniformly distributed $\{1, \dots, m+1\}$.

In our application, we observe a time series of forecasts and outcomes, $(\hat{F}_{t,h}, \pi_{t+h})_{t \in \mathbb{N}}$. Here, $\hat{F}_{t,h}$ is a forecast made at time $t$ for the realization $\pi_{t+h}$, where $h \geq 1$ is a fixed forecast horizon. 

Following \citet[Definition 3.5]{arnold_sequentially_2023}, we say that 
the sequence of probabilistic forecasts $(\hat{F}_{t,h})_{t \in \mathbb{N}}$ is \emph{probabilistically calibrated at horizon $h$} if the resulting sequence of PITs, $Z_t = Z_{\hat{F}_{t,h}}(\pi_{t+h})$, satisfies
    \[
    \mathcal{L}(Z_t \mid Z_j, 0 \le j \le t-h) = \text{UNIF}(0,1), \quad \text{for all  $t \in \mathbb{N}$},
    \]
    where the left-hand side denotes the conditional distribution of $Z_t$ given $Z_j$, $0 \le j \le t-h$. A sequence of ensemble forecasts $(\mathcal{X}_{t})_{t \in \mathbb{N}}$ of size $m$ is \emph{rank calibrated at horizon $h$} if the resulting sequence of ranks, $R_t = \text{rank}_{\mathcal{X}_{t}}(Y_{t+h})$, satisfies
    \[
    \mathcal{L}(R_t \mid R_j, 0 \le j \le t-h) = \text{UNIF}(\{1, \dots, m+1\}), \quad \text{for all $t \in \mathbb{N}$}.
    \]
When $t \le h$, the conditioning set is empty, so the statements are unconditional. 

For the one-step-ahead case where $h=1$, the conditional distribution of each PIT, $Z_t$, is the standard uniform distribution, irrespective of the values of the previous PITs. This is precisely the condition for the sequences of PITs $(Z_t)$ and ranks $(R_t)$ to be independent and identically distributed (i.i.d.). This connects the sequential framework directly to the classical notion of calibration as formulated by \citet{diebold_evaluating_1998}, which requires an i.i.d.~uniform sequence of PITs. For forecast horizons $h > 1$, the definition is more flexible. The conditioning is on PITs that are at least $h$ time steps in the past, which allows for potential dependence between test statistics that are less than $h$ time steps apart. This reflects the overlapping information content of multi-step-ahead forecasts.

Probabilistic calibration generally does not imply that the predictive distribution $\hat{F}_{t,h}$ matches the true conditional distribution of the outcome, i.e., $\mathcal{L}(Y_{t+h}|\mathcal{F}_t) = \hat{F}_{t,h}$ \citep{gneiting_combining_2013}. If the latter was the case, we would call the forecast \emph{ideal}. We focus on testing for probabilistic calibration, as it is more practical and a necessary condition for ideal forecasts. If a sequence of forecasts fails this test, it cannot be ideal. As noted by \citet{arnold_sequentially_2023}, probabilistically calibrated forecasts are generally not ideal when forecasts use side information in addition to the target variable's own history. The forecasting methods outlined in Section~\ref{sec:forecast_design} use a wide range of macroeconomic predictors and thus fall into this category.

\subsection{Anytime-valid inference on calibration}
\label{sec:evalsforcalibration}

\subsubsection{Preliminaries on e-values}
E-values allow to draw inference continuously over time without compromising type~I error guarantees. They have attracted great interest in the statistical literature in recent years for several reasons including their favorable behavior under optional stopping or continuation of experiments, and since they are easy to combine in contrast to p-values \citep{ramdas_game-theoretic_2023}. An e-value is a non-negative random variable with expectation less or equal to one under the null hypothesis. By Markov's inequality, the reciprocal of an e-value is a (typically conservative) p-value. Therefore, e-values reject the null hypothesis when they are large, so they can be interpreted as the amount of evidence against the null hypothesis. Indeed, \citet{shafer_testing_2021} nicely illustrate how an e-value can be seen as a bet with unit capital against the null hypothesis. If the null hypothesis is true, one can never make money on average.

Anytime-valid inference on calibration is based on sequential e-values. That is, we consider a sequence of adapted, non-negative random variables $(E_t)_{t \in \mathbb{N}}$ that satisfy
\[
\mathbb{E}(E_t | \mathcal{F}_{t-h}) \le 1 \quad  \text{ and for all  $t \in \mathbb{N}$ under calibration}.
\]
We call $E_t$ a \emph{sequential e-value} for calibration at lag $h$. For $t \le h$, the conditional expectation is understood unconditionally \citep[Definition 4.2]{arnold_sequentially_2023}.

For lag $h=1$, we construct a process by taking the cumulative product:
\begin{equation}
e_t = \prod^t_{i=1} E_i \text{ for } t \in \mathbb{N}. \label{tag:illustration}
\end{equation}
By the law of iterated expectations, the process $(e_t)_{t \in \mathbb{N}}$ is a non-negative supermartingale with expectation less or equal to one under the null hypothesis of calibration. Such a process is called a test supermartingale for the null hypothesis of calibration. Ville's inequality \citep{ville_etude_1939}, which holds for any test supermartingale, states that, under the null hypothesis, the probability of the process $(e_t)_{t \in \mathbb{N}}$ ever crossing the threshold $1/\alpha$ is at most $\alpha$:
\[
\mathbb{P} \left( \exists\, t \in \mathbb{N}: e_t \ge \frac{1}{\alpha} \right) \le \alpha.
\]
This inequality justifies anytime-valid inference. We can monitor $e_t$ continuously for $t \ge 1$ and stop the first time it crosses $1/\alpha$. The type-I error of this procedure is guaranteed to be no more than $\alpha$. Hence, every test supermartingale defines a sequential hypothesis test $\phi_t = \mathbf{1}_{\{e_t \ge 1/\alpha\}}$.

For longer forecast horizons ($h > 1$), the simple product as at \eqref{tag:illustration} does not allow for anytime-valid inference. 

\begin{prop}\citep[Prop. 4.3]{arnold_sequentially_2023}\label{prop:combine_seq_evalue}
Let $(E_t)_{t \in \mathbb{N}}$ be a sequence of sequential e-values for $\mathcal{H}_0 \subset \mathcal{P}$ at lag $h$, adapted to the filtration $(\mathcal{F}_t)_{t \in \mathbb{N}}$. For any time $t \ge h+1$, define the index sets $I_k(t) = \{k+hs : s = 0, \dots, \lfloor(t-k)/h\rfloor\}$. The aggregated value
\begin{equation}
e_t = \frac{1}{h} \sum_{k=1}^h \prod_{l \in I_k(t)} E_l \label{tag:combine_evals}
\end{equation}
is an $\mathcal{F}_t$-measurable e-value for $\mathcal{H}_0$. The process $(e_t)_{t \in \mathbb{N}}$ satisfies
\[
\mathbb{E}_P(e_{\tau+h-1}) \le 1, \quad \text{for any stopping time } \tau \text{ and any } P \in \mathcal{H}_0.
\]
\end{prop}

The properties of the process $(e_t)_{t \in \mathbb{N}}$ from Proposition~\ref{prop:combine_seq_evalue} depend on the lag $h$. For the one-step-ahead case where $h=1$, the process simplifies to the product at \eqref{tag:illustration}. In contrast, when $h > 1$, the process $(e_t)_{t \in \mathbb{N}}$ is generally not a test supermartingale. This is because $e_t$ is an average of $h$ distinct sub-processes, $M^{[k]} = (\prod_{l \in I_k(t)} E_l)_{t \in \mathbb{N}}$. Each sub-process $M^{[k]}$ is a supermartingale, but only with respect to its own specific, lagged filtration. It is not a supermartingale with respect to the underlying filtration $(\mathcal{F}_t)_{t \in \mathbb{N}}$. An average of processes that are supermartingales relative to different filtrations is not guaranteed to be a supermartingale itself. As a result, we cannot apply Ville's inequality directly.

There are two strategies to address this issue for $h>1$. The first strategy constructs a valid level-$\alpha$ test by modifying the stopping rule. As Ville's inequality still applies to each sub-process $M^{[k]}$ individually, we can track the supremum of each sub-process over time. We average these $h$ supremum processes into a single process, $\frac{1}{h} \sum_{k=1}^h \sup_{s \le t} \prod_{l \in I_k(s)} E_l$, and apply a penalty factor of $e \log(h)$, where $e$ is Euler's number. The modified rule is:
\begin{equation}
\tau_{\alpha,h} = \inf\left\{ t \in \mathbb{N} : \frac{1}{h e \log(h)} \sum_{k=1}^h \sup_{s \le t} \prod_{l \in I_k(s)} E_l \ge 1/\alpha \right\}.
\label{tag:mod_stoppingrule}
\end{equation}
This stopping time ensures that the probability of a false rejection is no more than $\alpha$, i.e., $\mathbb{P}_{\mathcal{H}_0}(\tau_{\alpha,h} < \infty) \le \alpha$ \citep{arnold_sequentially_2023}.

A second strategy, employed by \citet{henzi_valid_2022}, redefines the constituent e-values $E_t$. This approach creates a new process $(\tilde{e}_t)_{t \in \mathbb{N}}$ that is guaranteed to be a test supermartingale. However, the method can be very conservative. It works by multiplying each e-value by a correction factor that depends on the minimum possible values of future betting functions. If any future function can produce a value close to zero, the correction factor becomes very small. This severely dampens the process and reduces the power of the test. 

The practical challenge is to construct e-values that are powerful enough to detect miscalibration. We construct the e-value for each new observation as a likelihood ratio. This ratio compares the simple uniform null of a uniform distribution that results under calibration with an alternative distribution from the composite family, with parameters estimated from past data.

\subsubsection{E-values for the continuous uniform distribution}\label{ssec:cont_evals}
To test for probabilistic calibration, we assess if a sequence of PITs, $(z_t)_{t \in \mathbb{N}} \subseteq [0,1]$, consists of i.i.d.~draws from the standard uniform distribution. The null hypothesis is therefore the simple hypothesis $\mathcal{H}_0 := \{\text{UNIF}(0,1)\}$. For the alternative hypothesis, $\mathcal{H}_1$, we choose the family of beta distributions. This family can flexibly capture common forms of miscalibration, including bias (skewness) and dispersion errors (U-shaped or inverse-U-shaped distributions). Let $P_{\alpha, \beta}$ denote a beta distribution parametrized by \mbox{$\Theta = \{ (\alpha, \beta) \in \mathbb{R}^2 \mid \alpha > 0, \beta > 0 \}$}. The beta family nests the null hypothesis, since $P_{1,1}$ is the standard uniform distribution.

A fixed alternative, like a single beta distribution, is impractical because the nature of any miscalibration is unknown and can change over time. We therefore adopt the sequential, adaptive strategy from \citet{arnold_sequentially_2023}. This strategy uses the history of observations to inform the choice of parameters $\alpha$ and $\beta$ for the next test. This procedure constitutes a betting strategy: at each step, we bet on the alternative that best explains all data observed so far.

The fundamental component of this strategy is the likelihood ratio for a single observation $z \in [0,1]$. This ratio tests the simple null hypothesis that $z$ is drawn from $P_{1,1}$ against a simple alternative $P_{(\alpha, \beta)}$ for some $(\alpha, \beta) \neq (1,1)$. The likelihood ratio is:
\begin{equation}
    E^{\alpha,\beta}(z) = \frac{1}{B(\alpha, \beta)} z^{\alpha-1}(1-z)^{\beta-1}, \label{tag:betaevalue}
\end{equation}
where $B(\cdot, \cdot)$ is the beta function. By construction, $E^{\alpha,\beta}(z)$ is a valid e-value. In our sequential procedure, we use past data to select the alternative for the next observation. At each time $t \geq 2$, we compute the MLE of the beta parameters, $(\hat{\alpha}_{t}, \hat{\beta}_{t})$, using all previous observations $z_1, \dots, z_{t}$:
\begin{equation}
(\hat{\alpha}_t, \hat{\beta}_t) = \operatorname*{arg\,max}_{(\alpha,\beta) \in \Theta} \sum_{i=1}^t \log(p_{(\alpha, \beta)}(z_i)), \label{tag:mle}
\end{equation}
where $p_{(\alpha, \beta)}$ is the density of $P_{(\alpha, \beta)}$. We then use these parameters, which represent the data-driven estimate of the alternative, to form the e-value for the next observation, $z_{t+1}$:
\[
E_{t+1} = E^{\hat{\alpha}_{t},\hat{\beta}_{t}}(z_{t+1}).
\]
Setting $E_1 = E_2 = 1$, the procedure generates a sequence of e-values $(E_t)_{t \in \mathbb{N}}$. For a lag of $h=1$, the cumulative product of this sequence forms a test supermartingale, as in (\ref{tag:illustration}).

For sequential e-values at lag $h>1$, we must adapt the procedure to satisfy Definition~4.2 from \citet{arnold_sequentially_2023}. We perform the parameter estimation separately on $h$ disjoint subsequences of the data. For each $k \in \{1, \dots, h\}$, the $k$-th subsequence is defined by the indices $\{k + hs \mid s = 0,1,\dots\}$. For any time $t \ge 2h$, we compute $h$ separate MLEs on these subsequences:
\begin{equation}
(\hat{\alpha}_t^k, \hat{\beta}_t^k) = \operatorname*{arg\,max}_{(\alpha,\beta) \in \Theta} \sum_{s: k+hs \le t} \log(p_{(\alpha,\beta)}(z_{k+hs})), \quad k=1,\dots,h. \label{tag:laggedmle}
\end{equation}
The e-value for a new observation then uses the most recent lagged MLE from its corresponding subsequence. We initialize $E_1 = \dots = E_{2h} = 1$, and for $t = h,h+1, \dots$, we compute the sequential e-value as:
\[
E_{k+th} = E^{\hat{\alpha}_{t}^k,\hat{\beta}_{t}^k}(z_{k+th}), \quad k=1,\dots,h.
\]
The resulting sequence $(E_t)_{t\in\mathbb{N}}$ constitutes sequential e-values at lag $h$ for the uniform null hypothesis that $z_t \sim \text{UNIF}(0,1)$, conditional on $z_1, \dots, z_{t-h}$. As described in Proposition~\ref{prop:combine_seq_evalue}, we then combine these e-values using the formula in (\ref{tag:combine_evals}) to form an aggregated e-value at each time step.

In practice, we must address two potential pitfalls. First, the continuous uniform null assigns zero probability to the boundary points $\{0,1\}$, but empirical PITs can take these values. An observation of 0 or 1 can cause the MLE in (\ref{tag:mle}) to diverge. For the beta e-value in (\ref{tag:betaevalue}), this causes the likelihood ratio to become infinite or exactly zero. An e-value of zero is fatal for the cumulative product, as it permanently reduces the process to zero. In the betting interpretation of \citet{shafer_testing_2021}, this corresponds to losing all capital. We solve this problem by ignoring any observations in $\{0,1\}$ during parameter estimation and e-value calculation. This is a valid strategy because it does not alter the e-value's expectation under the null. As \citet{arnold_sequentially_2023} argue, this omission has little effect if boundary observations are rare; if they are frequent, the null hypothesis of a continuous uniform distribution is clearly false. Second, to ensure parameter estimates are stable, we begin the sequential estimation only after a warm-up period. Following \citet{arnold_sequentially_2023}, we set the first $n_0=10$ e-values to 1. This provides a buffer beyond the absolute minimum of two observations required to compute the MLE for the two-parameter beta distribution.

\subsubsection{E-values for the discrete uniform distribution}\label{ssec:discrete_evals}
To test for rank calibration, we assess if a sequence of ensemble ranks, $(r_t)_{t \in \mathbb{N}} \subseteq \{1, \dots, m\}$, is i.i.d.~uniform. For $m \geq 1$, the null hypothesis is therefore the simple hypothesis that $\mathcal{H}_0 := \{\text{UNIF}(\{1, \dots, m\})\}$. For the alternative hypothesis, $\mathcal{H}_1$, we use the family of beta-binomial distributions, parametrized by \mbox{$\Theta = \{ (\alpha, \beta) \in \mathbb{R}^2 \mid \alpha > 0, \beta > 0 \}$}. This choice is the discrete analogue of the beta distribution and can flexibly model common forms of miscalibration, like bias and dispersion errors in the rank histogram.

We construct the e-value using the same adaptive strategy as in the continuous case. The e-value for a single rank observation $r \in \{1, \dots, m\}$ is the likelihood ration of a specific alternative $P_{(\alpha, \beta)}$ to the uniform null:
\begin{equation}
E^{\alpha,\beta}(r) = \frac{p_{(\alpha, \beta)}(r)}{p_0(r)} = m \cdot p_{(\alpha, \beta)}(r), \label{tag:betabinomialevalue}
\end{equation}
where $p_0(r) = 1/m$ is the probability mass function (PMF) of the discrete uniform distribution and $p_{(\alpha, \beta)}(r)$ is the PMF of the beta-binomial distribution:
\[
p_{(\alpha, \beta)}(r) = \binom{m-1}{r-1} \frac{B(\alpha + r - 1, \beta + m - r)}{B(\alpha, \beta)}.
\]
By construction, the likelihood ratio in Equation~(\ref{tag:betabinomialevalue}) is an e-value for testing the uniform null against a simple beta-binomial alternative $P_{(\alpha, \beta)}$. We implement this test using a sequential procedure. For each new rank $r_{t+1}$ with $t \ge 2$, we first compute the MLE of the beta-binomial parameters, $(\hat{\alpha}_{t}, \hat{\beta}_{t})$, using all past ranks $r_1, \dots, r_{t}$:
\[
(\hat{\alpha}_t, \hat{\beta}_t) = \operatorname*{arg\,max}_{(\alpha,\beta) \in \Theta} \sum_{i=1}^t \log(p_{(\alpha, \beta)}(r_i)).
\]
We then form the e-value for the next rank $r_{t+1}$ using these estimated parameters:
\[
E_{t+1} = E^{\hat{\alpha}_{t},\hat{\beta}_{t}}(r_{t+1}).
\]
We initialize the sequence by setting $E_1 = E_2 = 1$ to have a sufficient sample for the first MLE. For a forecast horizon of $h=1$, the cumulative product of the resulting sequence $(E_t)_{t \in \mathbb{N}}$ forms a test supermartingale.

For sequential e-values at lag $h>1$, we perform the parameter estimation separately on the $h$ subsequences of the data, as in the continuous case. However, the beta-binomial MLE can be unstable with few observations, a problem exacerbated by partitioning the data. To ensure robust estimates, we again follow the recommendation of \citet{arnold_sequentially_2023} and introduce a warm-up period for each subsequence. Specifically, we set the first $n_0=10$ e-values for each of the $h$ subsequences to 1 before beginning their MLE procedure. After the warm-up, we compute the MLE for each $k$-th subsequence as in Equation~(\ref{tag:laggedmle}). The e-value for a subsequent observation is then calculated with the lagged estimates from its corresponding subsequence:
\[
E_{k+th} = E^{\hat{\alpha}_{t}^k,\hat{\beta}_{t}^k}(r_{k+th}), \quad k=1,\dots,h.
\]
The resulting sequence $(E_t)_{t\in\mathbb{N}}$ constitutes sequential e-values at lag $h$ for the discrete uniform null hypothesis, which we then combine using the formula in~(\ref{tag:combine_evals}).

\subsection{Implementation of sequential calibration tests}
\label{sec:application}
This section details how we apply the theoretical framework from Section~\ref{sec:evalsforcalibration}. We use this framework to evaluate the probabilistic inflation forecasts generated by the models in Section~\ref{sec:forecast_design}. Our evaluation is implemented in \texttt{R} using the \texttt{epit} package for the e-value computations~(\url{https://github.com/AlexanderHenzi/epit}). Our analysis simulates a realistic, real-time monitoring scenario, providing a dynamic assessment of forecast calibration. We structure the analysis by iterating through each combination of geographic region (United States, Euro Area, Switzerland), forecast horizon (in months, $h \in \{1, 3, 6, 12\}$), and forecasting model. To ensure a fair comparison, the evaluation period for each region begins only when all models provide valid, non-missing forecasts\footnote{For example, the series $(\hat{F}_{t,h}, \pi_t)_t$ is one month shorter for models requiring differencing of the time series (BVAR, DFM).}. 

We group the models into three types. For the majority of models that produce a Gaussian predictive distribution -- the Gaussian baselines, ARIMA, BVAR, and DFM -- we test calibration using the PIT. We compute the PIT for each forecast by evaluating the predicted Gaussian CDF at the realized inflation rate. For the probabilistic no-change model, which generates a non-parametric ensemble of $m$ recent inflation values, we assess calibration using ranks. We calculate the rank of the true outcome relative to the ensemble members. Finally, the distributional random forest produces a non-parametric distribution as a set of weighted samples, which results in a discontinuous CDF. Therefore, we compute a randomized PIT for each DRF forecast, as detailed in Section~\ref{sec:notionscalibration}.

With these sequences of PITs and ranks, we sequentially test the null hypothesis of calibration ($\mathcal{H}_0: z_t \sim \text{UNIF}(0,1)$ or $\mathcal{H}_0: r_t \sim \text{UNIF}\{1, \dots, m\}$). We employ the adaptive betting strategy from Section~\ref{sec:evalsforcalibration}. This requires choosing a family of alternative distributions against which to test the uniform null. For PITs, we use the beta distribution family; for ranks, we use the beta-binomial family. While \citet{arnold_sequentially_2023} also explore a kernel density method for the alternative, they find that tests based on the beta family are more powerful in small samples. We therefore adopt their beta-based approach. Finally, to handle PIT values that are exactly 0 or 1, we follow \citet{arnold_sequentially_2023} and exclude these boundary observations from the parameter estimation step, but we record their frequency for each model and horizon.

The method for aggregating the sequential e-values depends on the forecast horizon $h$. For one-step-ahead forecasts ($h=1$), the underlying information sets are non-overlapping. The cumulative product of the sequential e-values is therefore a test supermartingale. This allows us to use the standard stopping rule for calibration testing, as described in the paragraph following Proposition~\ref{prop:combine_seq_evalue}. In contrast, for multi-step-ahead forecasts ($h>1$), the information sets overlap. To handle this, we follow the U-statistics approach from \citet{arnold_sequentially_2023}, see Proposition \ref{prop:combine_seq_evalue}. We split the sequence of test statistics into $h$ non-overlapping subsequences, calculate a separate e-value sequence for each, and then aggregate them into a single value at each time step. The resulting sequence of aggregated e-values is not a test supermartingale. Therefore, to conduct a formal test, we must apply the less powerful, modified stopping rule from~(\ref{tag:mod_stoppingrule}).

The cumulative e-values for each model and forecast horizon are the primary output of our analysis. The e-values track how evidence against the null hypothesis builds up over time. For $h=1$, we treat this process as a statistically anytime-valid test. We reject $\mathcal{H}_0$ once it crosses a pre-specified threshold, like $1/\alpha = 100$ for a significance level of $\alpha = 0.01$.\footnote{In traditional scientific hypothesis testing, there are (discipline-specific) conventions on thresholds for p-values, typically 0.05 or 0.01. There is not yet a consensus on thresholds for e-values. For our purposes, we regard an e-value of around 100 as sufficient for a clear rejection. We regard a p-value of 0.01 as a clear rejection as well, but we still report p-values between 0.01 and 0.1. For a more thorough discussion of the comparison between e-values and p-values, see \citet[Sec. 2.7]{ramdas_hypothesis_2025}.} We also use the path of the process to identify when miscalibration appears within the evaluation period. For $h > 1$, we can formally reject the null hypothesis of calibration by applying the modified stopping rule and its corresponding threshold defined in \eqref{tag:mod_stoppingrule}. To provide a complete picture, we display these formal rejection thresholds alongside the supremum process in our graphical diagnostics. Beyond formal binary rejections, a major benefit of the e-value-based methodology is the ability to monitor the tests sequentially over time to see exactly when miscalibration develops. Sustained increases in the e-values provide valuable insights into accumulating evidence against forecast calibration, even during periods where the formal threshold has not yet been crossed.

The cumulative e-values show when our forecasts are miscalibrated, but they do not show how. We therefore complement them with graphical tools. A common static tool is the PIT histogram, which provides end-of-sample information about the type of miscalibration. Following \citet{arnold_sequentially_2023}, we go further and exploit the e-value procedure itself. The e-value construction specifies an alternative density, determined by the data at each time point. We plot these estimated densities over time to see in what way the forecasts are misspecified. In practice, we use the beta densities that enter the e-value construction. We also add non-parametric, boundary-corrected kernel density estimates. This additional method investigates if the beta specification captures the main features of the empirical PIT density. To compare the e-value-based test with a traditional static test, we compute the one-sample Kolmogorov-Smirnov (KS) test to evaluate the uniformity of the PITs. We apply the KS test to the full sample of PITs for each model except the probabilistic no-change forecast (\url{https://stat.ethz.ch/R-manual/R-devel/library/stats/html/ks.test.html}). Our analysis, therefore, contrasts dynamic evidence from the e-values with static, end-of-sample evidence from a classical test. This approach lets us detect whether a model is miscalibrated overall and identify the periods in which the forecasts are not calibrated. These features illustrate the practical value of sequentially valid inference for continuous monitoring.

\section{Probabilistic inflation forecasting}
\label{sec:forecast_design}
This section presents the results of the empirical application, which focuses on an out-of-sample forecasting experiment for probabilistic inflation forecasts. We begin by briefly describing the datasets employed and outlining the overall forecast design.

\subsection{Data and forecast design}
\label{sec:data}
Our empirical analysis focuses on three economies: the United States (US), the Euro Area (EA), and Switzerland (CH). We take the US data from the FRED-MD database maintained by the Federal Reserve Bank of St. Louis \citep{mccracken_fred-md_2015}. For the EA, we rely on data from the Statistical Data Warehouse of the European Central Bank. For Switzerland, we compile publicly available data from the Swiss National Bank\citep{swiss_national_bank_snb_2025}\footnote{\url{https://data.snb.ch/en}}.

For each economic region, we construct a harmonized set of predictors that balances cross-country comparability with data availability. We retain only variables that are available at a monthly frequency and free of missing observations over the relevant sample period. Each predictor set comprises approximately 20 time series capturing monetary and credit aggregates, interest rate conditions, as well as consumer and producer price developments. A complete overview of the included series is provided in Appendix~\ref{app:data}.

\begin{figure}[t]
  \centering
  \includegraphics[width=\textwidth]{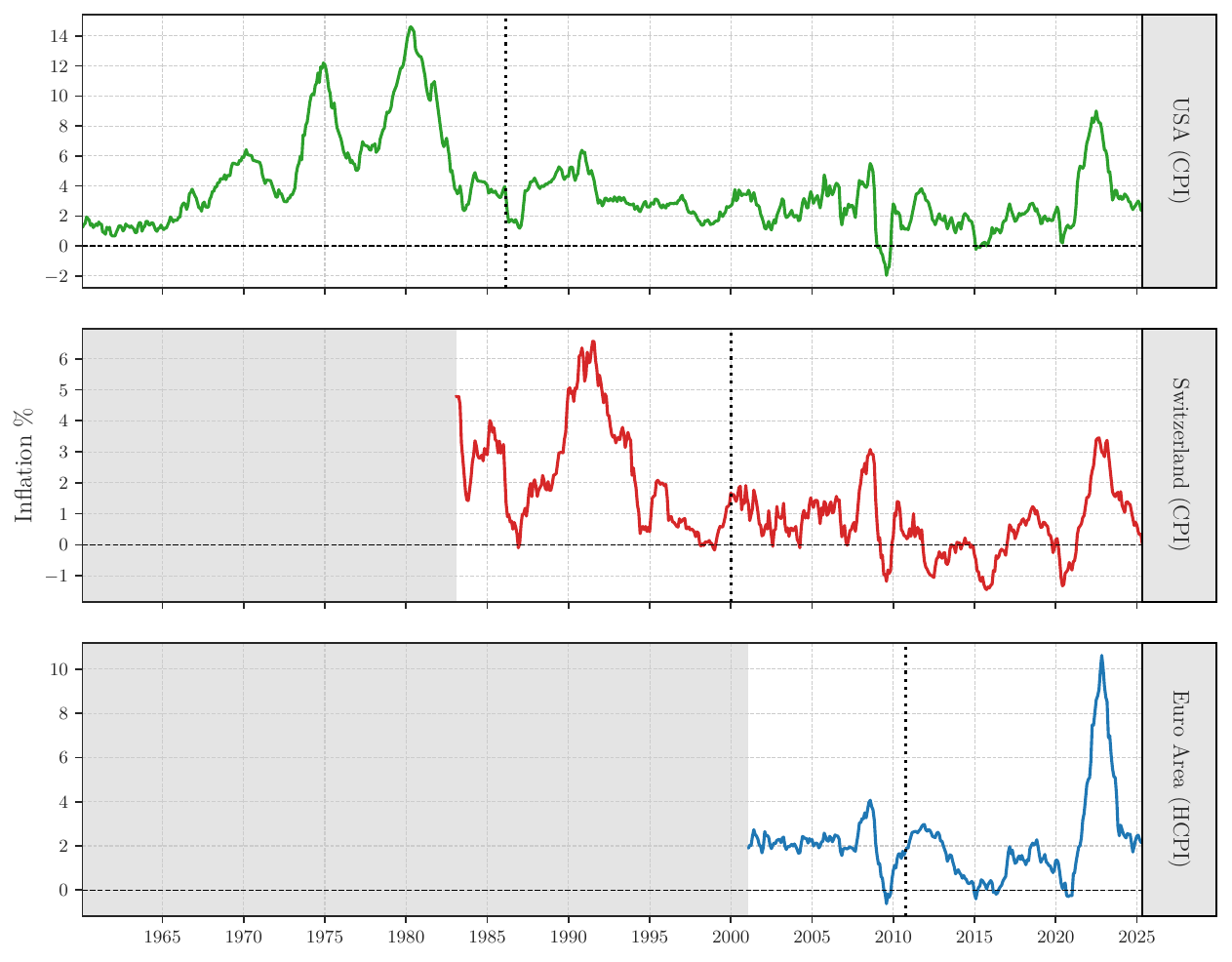}
  \vspace{-2.25em}
  \caption{Monthly year-over-year inflation rates for the three monetary regions. The vertical dotted line marks the end of the initial 40\% training period and the beginning of the out-of-sample evaluation window.}
  \label{fig:infl}
\end{figure}

The sample periods start at different dates but all end in April 2025. Our target variable is the year-over-year consumer price inflation rate, denoted by $\pi_t$, defined as
\begin{equation}
\pi_t \coloneq \frac{P_t}{P_{t-12}} - 1, \label{tag:inflation}
\end{equation}
where $P_t$ is the price index in month~$t$. We aim to forecast $\pi_{t+h}$ for forecast horizons $h \in \{1,3,6,12\}$. Forecasts at time~$t$ use only the information set $\mathcal{F}_t$, which contains all data available up to that month. This set includes past values of inflation $\{\pi_t, \pi_{t-1}, \dots\}$ and relevant macroeconomic predictors like interest rates and inflation components. The three datasets allow us to compare models and their calibration across economies with different inflation dynamics and sample sizes (see Figure~\ref{fig:infl}).

We mimic a realistic forecasting workflow as follows. We fix an initial training window at 40\% of each region's sample. At each evaluation month~$t$, we expand the training window to include all information up to $t$ and re-estimate the model parameters. We then issue the $h$-step predictive distribution $\hat{F}_{t,h}$ for inflation $\pi_{t+h}$. We update hyperparameters (like the number of lags) on a coarser schedule. We re-tune them whenever we observe at least a further 10\% of the full sample, that is, at 50\%, 60\%, $\dots$, and 90\% of each region's sample. Between these points, we keep hyperparameters fixed but still re-estimate model parameters each month using all data available at that time. Repeating this procedure yields a series ${(\hat{F}_{t,h}, \pi_{t+h})}_{t}$ for each horizon $h$.

The sample periods and sizes vary across the three regions. For the US, the data spans from January~1960 to April~2025 and contains 784 monthly observations. The out-of-sample evaluation begins after the initial training period ends in February~1986, leaving 470 monthly forecasts to evaluate. The Swiss dataset covers January~1983 to April~2025 (508 observations). The evaluation period starts after December~1999 and contains 305 forecasts. The Euro Area sample is the shortest. It runs from January~2001 to April~2025 (292 observations) and yields an evaluation window of 175 forecasts beginning after September~2010. Naturally, the evaluation windows are shorter for longer forecast horizons. Figure~\ref{fig:infl} plots the inflation series for each region, with a vertical line indicating the split between the initial training data and the evaluation period.

The target variable in each dataset is the year-over-year percentage change in the headline consumer price index: the Consumer Price Index for All Urban Consumers (CPIAUCSL) for the US, the Harmonized Index of Consumer Prices (HICP) for the EA, and the Swiss Consumer Price Index (CPI) for Switzerland. Some forecasting methods require stationary inputs, so we transform all predictor variables accordingly. We follow the transformations in \citet{mccracken_fred-md_2015} and the recommendations of each data provider. We report the exact transformations in Appendix~\ref{app:data}. To reduce outlier influence, we winsorize all stationary predictors at the 0.5th and 99.5th percentiles. We do not winsorize the inflation target series to preserve their original dynamics.

\subsection{Forecasting Models}

We consider a set of naive benchmark, standard time series, and machine learning models for computing the probabilistic inflation forecasts. Specifically, the set includes the naive last forecast, the rolling mean forecast, the probabilistic no-change forecast (PNC), the autoregressive integrated moving average model (ARIMA), Bayesian vector autoregressions with diffuse priors (BVAR (diffuse)), Minnesota priors (BVAR (Minnesota)), and Normal-Wishart priors (BVAR (NW)), the dynamic factor model (DFM), and the distributional random forest (DRF). All models are estimated recursively in an expanding-window framework and produce predictive distributions at horizons up to 12 months. Detailed descriptions of the models and their implementation are provided in Appendix~\ref{app:models}.

\begin{table}[!t]
  \centering
  \caption{Forecasting performance for the United States}
  \begin{threeparttable}
  \begin{tabular}{l cccccc}
\hline\hline
 & \multicolumn{3}{c}{$h=1$} & \multicolumn{3}{c}{$h=3$} \bigstrut[t]\\
\cmidrule(lr){2-4} \cmidrule(lr){5-7}
\textbf{Model} & \textbf{RMSE} & \textbf{MAE} & \textbf{CRPS} & \textbf{RMSE} & \textbf{MAE} & \textbf{CRPS} \bigstrut[b]\\
\hline
\multicolumn{7}{l}{\textit{Baseline Models}} \bigstrut[t]\\
Naive last & 0.398 & 0.276 & 0.206 & 0.865 & \textbf{0.575} & 0.441 \\
Rolling mean & 1.412 & 1.015 & 0.792 & 1.552 & 1.121 & 0.878 \\
PNC & 1.412 & 1.102 & 0.740 & 1.552 & 1.206 & 0.847 \\
ARIMA(1,1,0) & \textbf{0.362} & \textbf{0.259} & \textbf{0.191} & \textbf{0.849} & 0.577 & \textbf{0.436} \bigstrut[b]\\
\hline
\multicolumn{7}{l}{\textit{BVAR Models}} \bigstrut[t]\\
BVAR (diffuse) & 0.489 & 0.337 & 0.252 & 0.935 & 0.631 & 0.481 \\
BVAR (Minnesota) & 0.486 & 0.335 & 0.250 & 0.935 & 0.630 & 0.480 \\
BVAR (NW) & 0.483 & 0.332 & 0.249 & 0.920 & 0.617 & 0.471 \bigstrut[b]\\
\hline
\multicolumn{7}{l}{\textit{Multivariate \& ML Models}} \bigstrut[t]\\
DFM & 0.364 & 0.262 & 0.193 & 0.856 & 0.590 & 0.442 \\
DRF & 0.590 & 0.395 & 0.296 & 0.924 & 0.627 & 0.467 \bigstrut\\\hline\hline

 & \multicolumn{3}{c}{$h=6$} & \multicolumn{3}{c}{$h=12$} \bigstrut[t]\\
\cmidrule(lr){2-4} \cmidrule(lr){5-7}
\textbf{Model} & \textbf{RMSE} & \textbf{MAE} & \textbf{CRPS} & \textbf{RMSE} & \textbf{MAE} & \textbf{CRPS} \bigstrut[b]\\
\hline
\multicolumn{7}{l}{\textit{Baseline Models}} \bigstrut[t]\\
Naive last & 1.265 & 0.878 & 0.669 & 1.864 & 1.340 & 1.017 \\
Rolling mean & 1.716 & 1.246 & 0.983 & 1.907 & 1.419 & 1.125 \\
PNC & 1.716 & 1.316 & 0.975 & 1.907 & 1.489 & 1.135 \\
ARIMA(1,1,0) & 1.254 & 0.877 & 0.659 & 1.873 & 1.354 & 1.005 \bigstrut[b]\\
\hline
\multicolumn{7}{l}{\textit{BVAR Models}} \bigstrut[t]\\
BVAR (diffuse) & 1.317 & 0.921 & 0.715 & 1.933 & 1.376 & 1.103 \\
BVAR (Minnesota) & 1.318 & 0.926 & 0.715 & 1.931 & 1.376 & 1.100 \\
BVAR (NW) & 1.307 & 0.911 & 0.708 & 1.921 & 1.367 & 1.099 \bigstrut[b]\\
\hline
\multicolumn{7}{l}{\textit{Multivariate \& ML Models}} \bigstrut[t]\\
DFM & 1.266 & 0.888 & 0.665 & 1.911 & 1.374 & 1.033 \\
DRF & \textbf{1.173} & \textbf{0.785} & \textbf{0.575} & \textbf{1.341} & \textbf{0.864} & \textbf{0.620} \bigstrut\\\hline\hline
\end{tabular}
\vspace*{-0.5cm}
\begin{tablenotes}
    \singlespacing
    \item \leavevmode\kern-\scriptspace\kern-\labelsep 
    Note: The minimum value achieved for a given evaluation metric (RMSE, MAE, and CRPS) is highlighted in bold.
\end{tablenotes}
  \end{threeparttable}
  \label{tab:us_performance_summary}
\end{table}

\subsection{Out-of-Sample Forecasting Results}
\label{sec:forecast_summary}
In the following, we summarize the out-of-sample forecasting accuracy measured in terms of the RMSE, MAE, and CRPS, for the US and the Euro Area. The results are illustrated in Table~\ref{tab:us_performance_summary} and Table~\ref{tab:eu_performance_summary} in the Appendix. For both regions, the ARIMA(1,1,0) model consistently achieves the lowest errors across short horizons ($h = 1, 3$), closely followed by the DFM and the BVARs. The DRF performs competitively at longer horizons ($h = 6, 12$), but remains less accurate for near-term forecasts. The improved performance at longer forecasting horizons is likely attributable to the use of longer inflation lags in the DRF, which allows for a more effective representation of persistent dynamic behavior. The baseline models, especially the Gaussian no-change forecast (``naive last''), perform surprisingly well, confirming the persistence of monthly inflation. Overall, the relative ranking of the models is stable across the countries, and the results for the Switzerland (see Table~\ref{tab:ch_performance_summary} in the Appendix) mirror those for United States at slightly lower error levels. These findings provide an empirical foundation for the calibration evaluation in Section~\ref{sec:results}.

\section{Evaluation of probabilistic inflation forecasts}
\label{sec:results}

This section reports the main empirical findings. The sequential testing framework introduced in Section~\ref{sec:evaluation_theory} is employed to assess the probabilistic inflation forecasts produced by the models described in Section~\ref{sec:forecast_design}. The analysis primarily relies on cumulative e-values, as elaborated in Subsections~\ref{ssec:cont_evals} and~\ref{ssec:discrete_evals}, which facilitate a dynamic assessment of forecast calibration. To underscore the additional insights offered by the sequential perspective, these results are compared with those obtained from the static one-sample Kolmogorov–Smirnov (KS) test. The e-value approach allows not only for the detection of forecast miscalibration but also for the identification of its timing and nature.

The results are presented separately for each model class and focus on the United States and the Euro Area. Outcomes for the United States are shown in Figure~\ref{fig:us_advanced_short} and in Figures~\ref{fig:us_baseline_short}--\ref{fig:us_bvar_long} in the Appendix, while those for the Euro Area are reported in Figure~\ref{fig:eu_advanced_short} and Figures~\ref{fig:eu_baseline_short}--\ref{fig:eu_bvar_long} in the Appendix. Results for Switzerland are deferred to Appendix~\ref{chap:results_ch}. Each figure displays probability integral transform (PIT) histograms for the full evaluation sample using 20 bins in panel~(a)\footnote{For the PNC method, the normalized rank histogram is reported instead.} and the evolution of cumulative e-values over time in panel~(b). For $h > 1$, the dashed grey line represents the supremum process used for the modified stopping rule from \eqref{tag:mod_stoppingrule}. Panels~(a) and~(b) are organized by forecast horizon (columns) and model specification (rows). To examine changes in forecast misspecification over time, panel~(c) is included. This panel presents estimates of the alternative PIT density based on observations from the highlighted subsample (solid line) and from the preceding sample excluding this period (dotted line). The column headers indicate the corresponding highlighted time intervals. Two approaches are used to estimate the alternative density. First, a beta distribution is fitted, which also serves as the basis for constructing the e-values (blue lines). Second, a boundary-corrected kernel density estimator is applied, shown in black, following \citet{arnold_sequentially_2023}. 

\subsection{Results for the Baseline models}
\label{sec:baseline_results}
The baseline forecasts exhibit heterogeneous calibration properties. Some models maintain adequate calibration, mirroring their strong forecast accuracy. Others exhibit clear and significant miscalibration.

\begin{figure}[!t]
  \centering
  \includegraphics[width=1\textwidth]{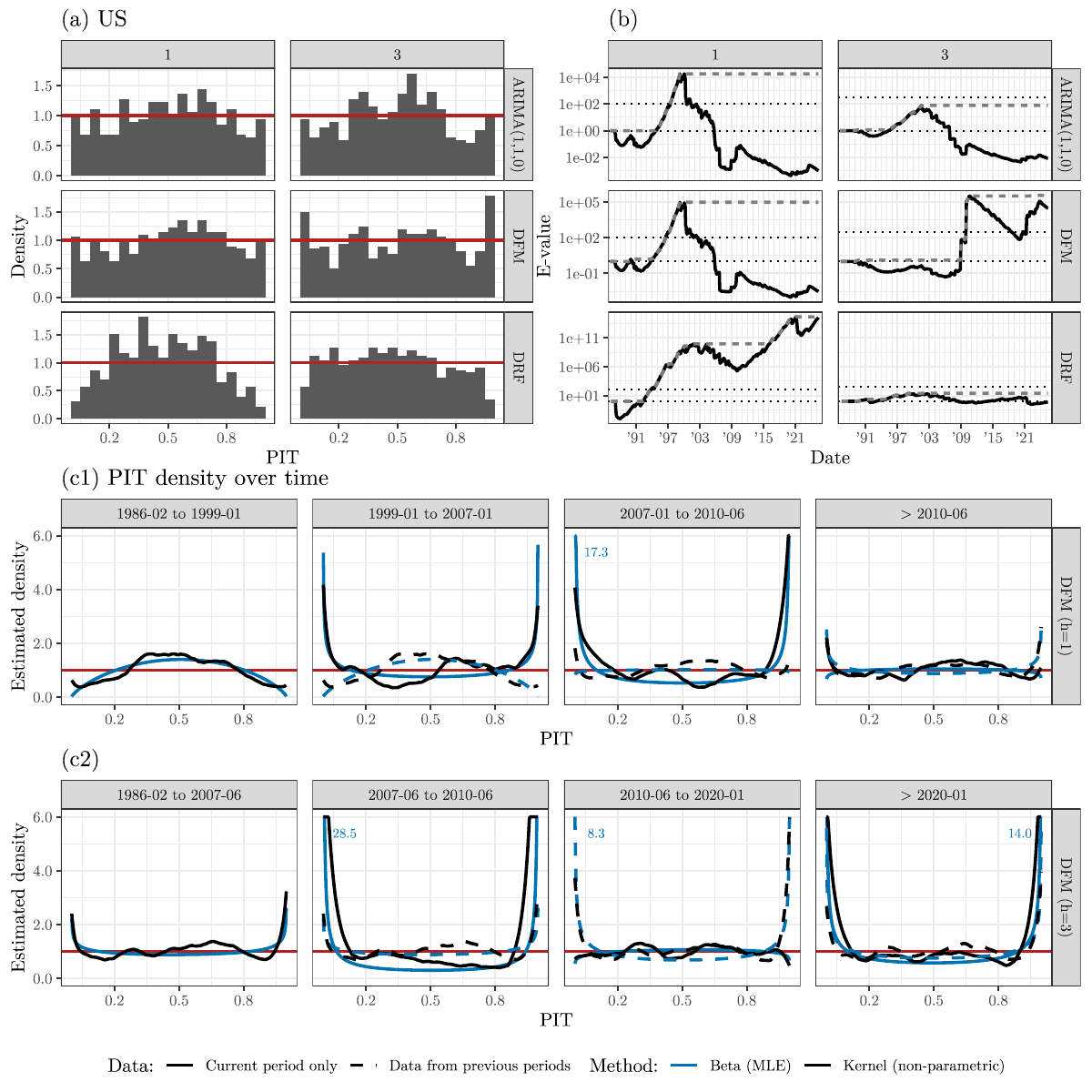}
    \vspace{-2.25em}
  \caption{Calibration diagnostics for ARIMA(1,1,0), DFM and DRF models at short horizons ($h=1,3$) for the United States. Panel (a) shows PIT histograms. Panel (b) shows cumulative e-values on a log scale; note the different y-scale across rows. The dotted horizontal lines indicate levels of 1 and the rejection thresholds. Panel (c) shows the evolution of the density estimates of the PIT; note that the y-axes are limited for better visibility, and the maximally attained value is given in the plot.}
  \label{fig:us_advanced_short}
\end{figure}

For the US at forecast horizon $h=1$, the ARIMA(1,1,0), rolling mean, and Gaussian no-change forecasts all exhibit notable miscalibration (Figures~\ref{fig:us_advanced_short} and~\ref{fig:us_baseline_short}). For the ARIMA(1,1,0) and Gaussian no-change approaches, the processes in Figures~\ref{fig:us_advanced_short} and~\ref{fig:us_baseline_short} show that most of the evidence against calibration accumulates during the 1990s. As shown in Table~\ref{tab:us_calibration_summary}, the static one-sample KS test misses these timing differences. It does not reject calibration for the ARIMA and Gaussian no-change forecasts for $h=1$, which is consistent with their approximately uniform full-sample PIT distributions.

At longer horizons, these forecasts behave differently, as shown in Figure~\ref{fig:us_advanced_long} in the Appendix. For the ARIMA(1,1,0) model, the evidence against calibration weakens as the horizon increases. Both the one-sample KS test and the cumulative e-value process support this conclusion (Table~\ref{tab:us_calibration_summary} in the Appendix). For the Gaussian no-change forecast, however, the cumulative e-values show diminishing power, while the KS test produces p-values in the range of $0.001-0.005$ (Figure~\ref{fig:us_baseline_long}).

The rolling mean forecast is clearly miscalibrated across all horizons. The static KS test rejects calibration with a p-value that is effectively zero, and the sequential analysis reaches the same conclusion: for $h=1$, the process reaches values of order $10^{34}$ (Table~\ref{tab:us_calibration_summary}). The e-values also reveal a distinct dynamic pattern. They accumulate evidence against calibration during relatively calm economic periods but drop sharply during the major crises in 2008-2009 and 2021-2023 (Figure~\ref{fig:us_baseline_short}).

\begin{figure}[!t]
  \centering
  \includegraphics[width=1\textwidth]{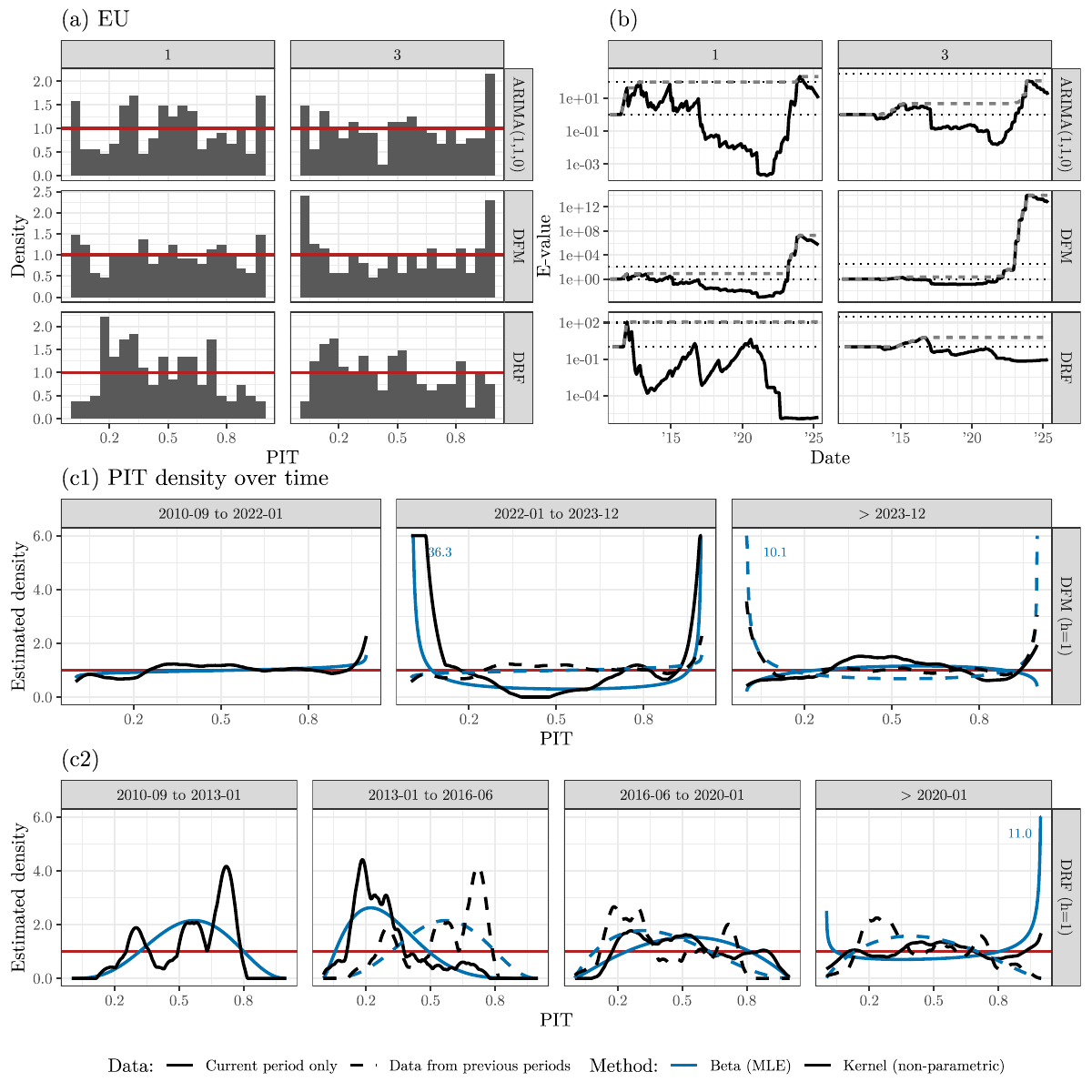}
    \vspace{-2.25em}
  \caption{Calibration diagnostics for ARIMA(1,1,0), DFM and DRF models at short horizons ($h=1,3$) for the Euro Area. Panel (a) shows PIT histograms. Panel (b) shows cumulative e-values on a log scale; note the different y-scale across rows. The dotted horizontal lines indicate levels of 1 and the rejection thresholds. Panel (c) shows the evolution of the density estimates of the PIT; note that the y-axes are limited for better visibility, and the maximally attained value is depicted in the plot.}
  \label{fig:eu_advanced_short}
\end{figure}

The Euro Area results illustrate the value of sequential testing in a relatively short sample with a pronounced regime shift (Figure~\ref{fig:infl}). For the one-month horizon, the processes for the baseline models rise sharply only at the end of the evaluation period, coinciding with the COVID-19 inflation shock (Figures~\ref{fig:eu_advanced_short} and~\ref{fig:eu_baseline_short}). This pattern is striking because the static one-sample KS test does not reject calibration for any of these methods at short horizons ($h \le 3$, see Table~\ref{tab:eu_calibration_summary}). The discrepancy suggests that the baseline forecasts suffered a transient but severe miscalibration during the pandemic shock. The sequential method detects this immediately, whereas the KS test, applied over the full sample, does not.

For longer horizons in the EA sample, the results are more complex. For the Gaussian no-change forecast, the cumulative e-values no longer provide evidence of miscalibration for $h \ge 3$. In contrast, for the ARIMA(1,1,0) model, the cumulative e-values continue to increase. They indicate miscalibration for $h=3$ and $h=6$ and reach values of order $10^{5}$ for $h=12$ (Figures~\ref{fig:eu_advanced_long} and~\ref{fig:eu_baseline_long}). The rolling mean method shows the familiar loss of power: its cumulative e-values no longer provide strong evidence of miscalibration (Figure~\ref{fig:eu_baseline_long}). The static KS test results at longer horizons are also difficult to interpret (Table~\ref{tab:eu_calibration_summary}). For example, for $h=6$ the one-sample KS test returns a small p-value only for the Gaussian no-change method ($p=0.04$), where the cumulative e-values do not grow. For $h=12$ it rejects calibration for the rolling mean forecast ($p=0.01$), where the e-values also remain flat.


\subsection{Results for the BVAR models}
\label{sec:bvar_results}
Despite their sophistication, the BVAR models are not calibrated across most regions and horizons. Their PIT histograms consistently show that the forecasts are overconfident. The choice of prior specification does not alter the overall dynamics of the miscalibration. Among the three priors, the Normal-Wishart prior produces the poorest results. Its e-values grow fastest and reach the highest peaks, providing the strongest evidence of miscalibration.

For the United States, the global financial crisis (GFC) of 2008–2009 is the main episode of BVAR miscalibration. Across all forecast horizons, the cumulative e-values increase sharply during this period and reach levels of order $10^{15}$ (for the Minnesota prior at $h=3$) up to $10^{36}$ (for the Normal–Wishart prior at $h=12$). The PIT histograms support this conclusion: they show clear overconfidence for $h=1$, which becomes more pronounced at longer horizons. Despite the clear results from the sequential test, the static KS test does not reject calibration for $h=1$ (Table~\ref{tab:us_calibration_summary}). The test supermartingale, in contrast, makes the miscalibration at this horizon visible (Figure~\ref{fig:us_bvar_short}). It also provides a timeline for when the BVARs' probabilistic calibration deteriorates, which is valuable for forecasters and policymakers who monitor models in practice.


For the EA, the cumulative e-values for all BVAR specifications grow extremely rapidly during the 2021-2022 inflation surge. For the three-month horizon ($h=3$), the Normal-Wishart prior BVAR's cumulative e-value exceeds the rejection threshold by mid-2021 (Figure~\ref{fig:eu_bvar_short}). This constitutes evidence against calibration. While slightly less extreme, the Minnesota and diffuse prior BVARs also reach high values over the same period. The PIT histograms in Figure~\ref{fig:eu_bvar_short} show that many realized inflation rates lie in the tails of the predictive densities. The speed at which the e-values reach these levels confirms that the BVARs did not adjust their uncertainty estimates in time for the regime change. The sequential analysis highlights a dynamic that the static KS test misses, particularly at short horizons. While the one-sample KS test does not reject calibration for $h=1$ (Table~\ref{tab:eu_calibration_summary}), the process indicates miscalibration as it unfolds. The KS test does eventually detect the issue, but only for longer horizons.

%
Across all priors and forecast horizons, the e-values tell a consistent story. They show no evidence against calibration during the first 20 years of stable inflation. They then exhibit a sharp, sudden spike during the COVID-19 shock (Figure~\ref{fig:eu_bvar_long}). The miscalibration during this shock overcomes the power loss of the sequential test at longer horizons. This finding exemplifies the advantage of using e-values in a realistic monitoring context: the BVARs' misspecification was evident as it happened.


\subsection{Results for the DFM}
\label{sec:dfm_results}
The DFM forecasts present a mixed picture. They are generally better calibrated than the BVARs, and their PIT histograms and e-values show properties similar to the ARIMA(1,1,0) model. The DFM results for the US provide further insights into the model's dynamic calibration. For the one-month horizon ($h=1$), the process shows an evolution similar to the ARIMA(1,1,0) model. It reaches an order of magnitude of $10^4$ during the 1990s before decreasing until the end of the evaluation period. On the other hand, the PIT histogram in panel~(a) of Figure~\ref{fig:us_advanced_short} looks approximately uniform. To investigate this in detail, we look at the evolution of the estimated alternative densities over selected time windows in panel~(c1) of Figure~\ref{fig:us_advanced_short}. We see from the inverse-U-shape in the first subplot that the initial increase is due to an overdispersion of the forecasts. Later, in the early 2000s, the forecasts were too narrow because the estimated alternative during that period was U-shaped. However, the process decreases during that time period. This decrease illustrates that the e-values only have power if the type of miscalibration aligns with the currently chosen alternative distribution. The decrease does not mean that the forecasts are better calibrated during this time period, but the misspecification merely changes direction. There is a short rise in the process during the GFC due to underdispersion of the forecasts, which subsides after the GFC. Now, the test supermartingale has power for this miscalibration because the dotted line (the current alternative) is also U-shaped. From 2010 onwards, the forecasts seem to depict uncertainty accurately. The static KS test, however, gives a p-value of $0.97$ against uniformity (Table~\ref{tab:us_calibration_summary}). Hence, this is an exemplary case of the usefulness of the sequential method, as it allows insights that are not possible through traditional methods.


At longer horizons, the PIT histograms begin to show clear signs of overconfidence, with values bunching at 0 and 1. Still, the static KS test only finds evidence against calibration for $h=12$ with a p-value of $0.02$ (Table~\ref{tab:us_calibration_summary}). The e-values for $h > 3$ continue the pattern seen for $h=3$: a period of calibration in the 1990s followed by a steep increase during the 2008-2009 crisis. However, the e-values again show a loss of power as the horizon lengthens. While the GFC shock is visible for $h=6$, the increase is less pronounced than for $h=3$ and does not provide the same level of immediate evidence. Instead, strong evidence of miscalibration at this horizon only accumulates later, during the COVID-19 shock in 2021 (Figure~\ref{fig:us_advanced_long}).

For the EA, the DFM's one-month-ahead forecasts initially appear calibrated, and in fact, better calibrated than the ARIMA model for much of the evaluation period. The PIT histogram for the DFM for $h=1$ is fairly uniform, and the one-sample KS test returns a p-value of $0.815$, indicating no grounds for rejection. However, the sequential test tells a different story. The process for the DFM remains near 1 for most of the sample, but begins to rise sharply in late 2022. It ultimately reaches a level of order $10^7$, decisively rejecting calibration for the one-month horizon. To check the type of miscalibration, we look at the evolution of the estimated alternative distribution in panel~(c1) of Figure~\ref{fig:eu_advanced_short}. The first subplot confirms that the DFM was probabilistically calibrated before the COVID-19 shock, because the alternative densities align well with the uniform line. By contrast, in 2022 and 2023, the estimated alternative is U-shaped. Therefore, prediction uncertainty was underestimated. After 2023, the process levels off and based on the mild inverse-U-shaped density estimate from this period, calibration was approximately appropriate. This pattern is robust across the longer horizons, where the cumulative e-values provide evidence of miscalibration, growing to orders of magnitude between $10^{12}$ and $10^{25}$ (Figure~\ref{fig:eu_advanced_long}). This detailed analysis provides valuable insight, especially for the shorter horizons, as the static KS test only begins to clearly reject the null hypothesis of calibration for $h \geq 6$ (Table~\ref{tab:eu_calibration_summary}).

\subsection{Results for the DRF}
\label{sec:drf_results}
The DRF displays yet another distinct calibration profile. As noted in Section~\ref{sec:forecast_summary}, the DRF's design matrix, which includes longer inflation lags, leads to higher forecasting accuracy at longer horizons. This characteristic generally extends to its calibration, where we see fewer issues as the horizon increases.

For the US, where the DRF benefits from a longer training period, calibration issues are most pronounced at the short horizon. For $h=1$, the cumulative e-value reaches a high level of order $10^{14}$, providing evidence against calibration (Figure~\ref{fig:us_advanced_short}). This finding is supported by low p-values from the static KS tests (essentially zero for $h=1$) and by the clearly inverse U-shaped PIT histograms (Table~\ref{tab:us_calibration_summary}). For longer horizons, the picture becomes more nuanced. The e-values no longer indicate evidence against calibration, and the PIT histograms appear less inverse U-shaped (Figure~\ref{fig:us_advanced_long}). The one-sample KS test, however, continues to suggest miscalibration for $h=3$ (p-value of $0.016$) and only fails to reject the null hypothesis for $h=12$ clearly (p-value of $0.183$).

For the EA, the DRF's one-month-ahead forecasts show early signs of miscalibration that appear to be transient. The PIT histogram for $h=1$ is inversely U-shaped, suggesting overdispersion. The test supermartingale reflects this dynamic. It rises initially to a value of 100, but subsequently drops below one (Figure~\ref{fig:eu_advanced_short}). The process then rises and falls repeatedly, remaining below 1 for the majority of the time. In contrast, the static KS test rejects calibration for $h=1$ with a p-value of $0.004$ (Table~\ref{tab:eu_calibration_summary}). 


For longer horizons ($h \geq 3$), the analysis is constrained by the very limited sample size, which makes definitive conclusions difficult. The cumulative e-values for these horizons do not provide any evidence against calibration. While the corresponding PIT histograms appear closer to uniform than for the one-month horizon, the picture from the one-sample KS test remains mixed. It produces p-values of $0.057$ and $0.098$ for horizons $h = 3$ and 6 but does not reject calibration for the longest horizon of $h=12$ with a p-value of $0.154$ (Figure~\ref{fig:eu_advanced_long}, Table~\ref{tab:eu_calibration_summary}).

\section{Conclusion}\label{sec:conclusion}
This paper investigates the calibration of probabilistic inflation forecasts for the United States, the Euro Area, and Switzerland, using a novel framework of sequentially valid inference to evaluate a range of forecasting models. By relying on e-values, we implement a dynamic assessment that reflects the realistic, sequential nature of macroeconomic forecasting and moves beyond traditional fixed-sample diagnostics. This allows us to identify not only whether a model is calibrated, but also when and under what conditions calibration deteriorates.

Our empirical results show that model complexity does not guarantee reliable uncertainty quantification. In fact, simpler models often provided the most trustworthy assessments of forecast uncertainty. The Gaussian no-change benchmark and the univariate AR models, were surprisingly well calibrated across many regions and horizons. By contrast, the Bayesian VAR models were systematically overconfident, with predictive distributions that were too narrow. Sequential testing was crucial in detecting this miscalibration, especially around the post-pandemic inflation surge in the Euro Area and the global financial crisis in the United States. The dynamic factor model performed better than the Bayesian VARs and was often close to the AR benchmark, though sequential diagnostics still revealed horizon-dependent miscalibration in the US forecasts. 

Beyond these model-specific findings, the paper demonstrates the practical value of sequentially valid testing for forecast evaluation in macroeconomics. Traditional tools such as the Kolmogorov-Smirnov test assume a fixed evaluation period, an assumption that is often violated in practice when forecasters monitor models continuously and revisit evaluation decisions over time. Such optional stopping undermines classical p-values. E-values address this problem by providing a statistically rigorous way to test for miscalibration at any time, as often as needed, without inflating false positive rates. This has clear implications for model governance. Forecast users, central banks, and financial institutions can use this framework to monitor calibration continuously and detect when a forecasting model becomes unreliable. A sharp and sustained rise in the e-values signals that the model is misspecified and should be re-estimated, re-specified, or replaced. In this way, sequential calibration monitoring can support more robust forecasting systems and, ultimately, better-informed policy and decision-making.

{\singlespacing
\addtocontents{toc}{\vspace{.5\baselineskip}}
\phantomsection
\addcontentsline{toc}{section}{\protect\numberline{}{Bibliography}}
\bibliography{myReferences,references}
}

\newpage

\setcounter{table}{0}
\setcounter{figure}{0}
\renewcommand\thefigure{\thesection.\arabic{figure}}
\renewcommand\thetable{\thesection.\arabic{table}}

\begin{center}
{\bf \Large Appendix}
\end{center}

\addtocontents{toc}{\vspace{.5\baselineskip}}
\appendix
\section{Data details}
\label{app:data}
This appendix provides details on the macroeconomic datasets used for the United States, the Euro Area, and Switzerland. To ensure comparability, we constructed a consistent set of predictors for each region, drawing from established macroeconomic databases.

For the Euro Area, we use the EA-MD-QD database compiled by \citet{barigozzi_ea-md-qd_2025}. From this dataset, we selected variables related to interest rates (indices 75-77), prices (indices 92-103), and monetary aggregates (indices 114-116). The dataset for the United States is the FRED-MD database from \citet{mccracken_fred-md_2015}. Our predictor set includes variables from group 5 (money and credit), group 6 (interest and exchange rates), and group 7 (prices). The data for Switzerland were primarily compiled from the SNB's data portal, supplemented with global variables such as crude oil prices from the FRED-MD database. To ensure the analysis reflects a consistent point in time for data availability, the vintages used are April 2025 for Switzerland, June 2025 for the Euro Area, and July 2025 for the United States.

Tables \ref{tab:us_data}, \ref{tab:eu_data}, and \ref{tab:ch_data} provide a comprehensive list of all variables used for each region, including their descriptions, sources, and the transformations applied to ensure stationarity for the relevant forecasting models.

\begin{table}[h]
\footnotesize
\centering
\begin{threeparttable}
\caption{US Dataset Description}
\label{tab:us_data}

\begin{tabular}{l l l l}
\hline\hline
\textbf{ID} & \textbf{Description} & \textbf{Unit} & \textbf{Transformation} \bigstrut \\
\hline
CPI & CPI All Items: CPIAUCSL & Percent & $\Delta x_t$ \bigstrut[t]\\
M1SL & M1 Money Stock & BLN\$ & $\Delta^2 \log(x_t)$ \\
M2SL & M2 Money Stock & BLN\$ & $\Delta^2 \log(x_t)$ \\
M2REAL & Real M2 Money Stock & BLN\$ & $\Delta^2 \log(x_t)$ \\
BUSLOANS & Commercial and Industrial Loans & BLN\$ & $\Delta^2 \log(x_t)$ \\
FEDFUNDS & Effective Federal Funds Rate & Percent & $\Delta x_t$ \\
TB3MS & 3-Month Treasury Bill & Percent & $\Delta x_t$ \\
TB6MS & 6-Month Treasury Bill & Percent & $\Delta x_t$ \\
GS1 & 1-Year Treasury Constant Maturity Rate & Percent & $\Delta x_t$ \\
GS5 & 5-Year Treasury Constant Maturity Rate & Percent & $\Delta x_t$ \\
GS10 & 10-Year Treasury Constant Maturity Rate & Percent & $\Delta x_t$ \\
PPICMM & PPI: Metals and metal products & 1982=100 & $\Delta^2 \log(x_t)$ \\
OILPRICEX & Crude Oil Prices & \$ per Barrel & $\Delta^2 \log(x_t)$ \\
CPIAPPSL & CPI: Apparel & 1982--84=100 & $100 \times \Delta \log(x_t)$ \\
CPITRNSL & CPI: Transportation & 1982--84=100 & $100 \times \Delta \log(x_t)$ \\
CPIMEDSL & CPI: Medical Care & 1982--84=100 & $100 \times \Delta \log(x_t)$ \\
CUSR0000SAC & CPI: Commodities & 1982--84=100 & $100 \times \Delta \log(x_t)$ \\
CUSR0000SAD & CPI: Durables & 1982--84=100 & $100 \times \Delta \log(x_t)$ \\
CUSR0000SAS & CPI: Services & 1982--84=100 & $100 \times \Delta \log(x_t)$ \\
PCEPI & Personal Consumption Expenditures Price Index & 2012=100 & $100 \times \Delta \log(x_t)$ \bigstrut[b]\\
\hline\hline
\end{tabular}
\begin{tablenotes}
  \item Notes: The data are collected from the FRED-MD dataset \citep{mccracken_fred-md_2015}. The operator $\Delta$ denotes first differencing, and all series are observed at a monthly frequency.
\end{tablenotes}

\end{threeparttable}
\end{table}

\clearpage

\begin{table}[h]
\footnotesize
\centering
\begin{threeparttable}
\caption{Euro Area Dataset Description}
\label{tab:eu_data}

\begin{tabular}{l l l l}
\hline\hline
\textbf{ID} & \textbf{Description} & \textbf{Unit} & \textbf{Transformation} \bigstrut \\
\hline
HCPI & HICP Overall index: HICPOV & Percent & $\Delta x_t$ \bigstrut[t]\\
IRT3M & 3-Month Interest Rates & Percent & $\Delta x_t$ \\
IRT6M & 6-Month Interest Rates & Percent & $\Delta x_t$ \\
LTIRT & Long-Term Interest Rates (EMU Criterion) & Percent & $\Delta x_t$ \\
PPICAG & Producer Price Index: Capital Goods & 2021=100 & $100 \times \Delta \log(x_t)$ \\
PPICOG & Producer Price Index: Consumer Goods & 2021=100 & $100 \times \Delta \log(x_t)$ \\
PPINDCOG & Producer Price Index: Non-Durable Consumer Goods & 2021=100 & $100 \times \Delta \log(x_t)$ \\
PPIDCOG & Producer Price Index: Durable Consumer Goods & 2021=100 & $100 \times \Delta \log(x_t)$ \\
PPIING & Producer Price Index: Intermediate Goods & 2021=100 & $100 \times \Delta \log(x_t)$ \\
PPINRG & Producer Price Index: Energy & 2021=100 & $100 \times \Delta \log(x_t)$ \\
HICPNEF & HICP: All Items excl. Energy \& Food & 2010=100 & $100 \times \Delta \log(x_t)$ \\
HICPG & HICP: Goods & 2010=100 & $100 \times \Delta \log(x_t)$ \\
HICPIN & HICP: Industrial Goods & 2010=100 & $100 \times \Delta \log(x_t)$ \\
HICPSV & HICP: Services & 2010=100 & $100 \times \Delta \log(x_t)$ \\
HICPNG & HICP: Energy & 2010=100 & $100 \times \Delta \log(x_t)$ \\
CURR & Money Stock: Currency in Circulation & MLN\,€ & $\Delta \log(x_t)$ \\
M1 & Money Stock: M1 & MLN\,€ & $\Delta \log(x_t)$ \\
M2 & Money Stock: M2 & MLN\,€ & $\Delta \log(x_t)$ \\
OILPRICEX & Crude Oil Prices & \$ per Barrel & $\Delta^2 \log(x_t)$ \bigstrut[b]\\
\hline\hline
\end{tabular}
\begin{tablenotes}
  \item Notes: The data are obtained from the Statistical Data Warehouse of the European Central Bank. The crude oil price is taken exclusively from the FRED-MD database. The operator $\Delta$ denotes first differencing, and all series are observed at a monthly frequency.
\end{tablenotes}

\end{threeparttable}
\end{table}

\clearpage

\begin{table}[h]
\footnotesize
\centering
\begin{threeparttable}
\caption{Switzerland Dataset Description}
\label{tab:ch_data}
\begin{tabular}{p{6cm} p{4.5cm} p{2.5cm} p{2.5cm}}
\hline\hline
\textbf{ID} & \textbf{Description} & \textbf{Unit} & \textbf{Transformation} \bigstrut \\
\hline
CPI & CPI Total & Percent & $\Delta x_t$ \bigstrut[t]\\
CPI\_GOODS & CPI: Goods & Dec 2020=100 & $100 \times \Delta \log(x_t)$ \\
CPI\_SERVICES & CPI: Services & Dec 2020=100 & $100 \times \Delta \log(x_t)$ \\
CPI\_HOUSING\_ENERGY & CPI: Housing and Energy & Dec 2020=100 & $100 \times \Delta \log(x_t)$ \\
CPI\_FOOD\_NONALCOHOL & CPI: Food and Non-alcoholic Beverages & Dec 2020=100 & $100 \times \Delta \log(x_t)$ \\
CPI\_TRANSPORT & CPI: Transport & Dec 2020=100 & $100 \times \Delta \log(x_t)$ \\
CPI\_HEALTH & CPI: Health & Dec 2020=100 & $100 \times \Delta \log(x_t)$ \\
CPI\_CLOTHING\_FOOTWEAR & CPI: Clothing and Footwear & Dec 2020=100 & $\Delta^2 \log(x_t)$ \\
CPI\_ALCOHOL\_TOBACCO & CPI: Alcoholic Beverages and Tobacco & Dec 2020=100 & $100 \times \Delta \log(x_t)$ \\
CPI\_HOUSEHOLD\_MAINTENANCE & CPI: Household Furniture and Maintenance & Dec 2020=100 & $100 \times \Delta \log(x_t)$ \\
CPI\_RESTAURANTS\_HOTELS & CPI: Restaurants and Hotels & Dec 2020=100 & $100 \times \Delta \log(x_t)$ \\
CPI\_RECREATION\_CULTURE & CPI: Recreation and Culture & Dec 2020=100 & $\Delta^2 \log(x_t)$ \\
CPI\_COMMUNICATIONS & CPI: Communications & Dec 2020=100 & $\Delta^2 \log(x_t)$ \\
CPI\_EDUCATION & CPI: Education & Dec 2020=100 & $100 \times \Delta \log(x_t)$ \\
M0 & Monetary Aggregate M0 Total & MLN CHF & $\Delta^2 \log(x_t)$ \\
PPI & Producer Price Index: Total & Dec 2020=100 & $100 \times \Delta \log(x_t)$ \\
IPI & Import Price Index: Total & Dec 2020=100 & $100 \times \Delta \log(x_t)$ \\
OILPRICEX & Crude Oil Prices & \$ per Barrel & $\Delta^2 \log(x_t)$ \bigstrut[b]\\
\hline\hline
\end{tabular}
\begin{tablenotes}
  \item Notes: The data are obtained from the Swiss National Bank. The crude oil price is taken exclusively from the FRED-MD database. The operator $\Delta$ denotes first differencing, and all series are observed at a monthly frequency.
\end{tablenotes}

\end{threeparttable}
\end{table}

\section{Further results}
\label{app:results}

\subsection{Further forecasting performance results}
\label{app:forecast_performance}

\begin{table}[!h]
  \centering
  \caption{Forecasting performance for the Euro Area}
   \begin{threeparttable}
  \begin{tabular}{l cccccc}
\hline\hline
 & \multicolumn{3}{c}{$h=1$} & \multicolumn{3}{c}{$h=3$} \bigstrut[t]\\
\cmidrule(lr){2-4} \cmidrule(lr){5-7}
\textbf{Model} & \textbf{RMSE} & \textbf{MAE} & \textbf{CRPS} & \textbf{RMSE} & \textbf{MAE} & \textbf{CRPS} \bigstrut[b]\\
\hline
\multicolumn{7}{l}{\textit{Baseline Models}} \bigstrut[t]\\
Naive last & 0.360 & 0.241 & 0.182 & 0.813 & 0.572 & 0.432 \\
Rolling mean & 2.026 & 1.376 & 1.038 & 2.275 & 1.560 & 1.169 \\
PNC & 2.026 & 1.369 & 0.980 & 2.275 & 1.556 & 1.163 \\
ARIMA(1,1,0) & \textbf{0.334} & \textbf{0.222} & \textbf{0.170} & \textbf{0.760} & 0.543 & \textbf{0.405} \bigstrut[b]\\
\hline
\multicolumn{7}{l}{\textit{BVAR Models}} \bigstrut[t]\\
BVAR (diffuse) & 0.431 & 0.289 & 0.224 & 1.065 & 0.714 & 0.568 \\
BVAR (Minnesota) & 0.426 & 0.285 & 0.219 & 1.039 & 0.693 & 0.550 \\
BVAR (NW) & 0.430 & 0.288 & 0.223 & 1.059 & 0.710 & 0.568 \bigstrut[b]\\
\hline
\multicolumn{7}{l}{\textit{Multivariate \& ML Models}} \bigstrut[t]\\
DFM & 0.342 & 0.225 & 0.170 & 0.791 & \textbf{0.543} & 0.417 \\
DRF & 1.295 & 0.712 & 0.508 & 1.492 & 0.910 & 0.639 \bigstrut\\\hline\hline

 & \multicolumn{3}{c}{$h=6$} & \multicolumn{3}{c}{$h=12$} \bigstrut[t]\\
\cmidrule(lr){2-4} \cmidrule(lr){5-7}
\textbf{Model} & \textbf{RMSE} & \textbf{MAE} & \textbf{CRPS} & \textbf{RMSE} & \textbf{MAE} & \textbf{CRPS} \bigstrut[b]\\
\hline
\multicolumn{7}{l}{\textit{Baseline Models}} \bigstrut[t]\\
Naive last & 1.403 & 0.934 & 0.728 & 2.430 & 1.610 & 1.240 \\
Rolling mean & 2.586 & 1.794 & 1.335 & 2.955 & 2.082 & 1.537 \\
PNC & 2.586 & 1.794 & 1.400 & 2.955 & 2.112 & 1.716 \\
ARIMA(1,1,0) & \textbf{1.335} & \textbf{0.894} & \textbf{0.699} & \textbf{2.385} & \textbf{1.583} & \textbf{1.253} \bigstrut[b]\\
\hline
\multicolumn{7}{l}{\textit{BVAR Models}} \bigstrut[t]\\
BVAR (diffuse) & 1.853 & 1.174 & 1.001 & 3.132 & 2.024 & 1.798 \\
BVAR (Minnesota) & 1.814 & 1.142 & 0.971 & 3.077 & 1.963 & 1.738 \\
BVAR (NW) & 1.841 & 1.168 & 1.001 & 3.118 & 2.018 & 1.800 \bigstrut[b]\\
\hline
\multicolumn{7}{l}{\textit{Multivariate \& ML Models}} \bigstrut[t]\\
DFM & 1.433 & 0.943 & 0.751 & 2.711 & 1.758 & 1.453 \\
DRF & 1.669 & 1.059 & 0.736 & \textbf{1.832} & \textbf{1.114} & \textbf{0.804} \bigstrut\\\hline\hline
\end{tabular}
\vspace*{-0.5cm}
\begin{tablenotes}
    \singlespacing
    \item \leavevmode\kern-\scriptspace\kern-\labelsep 
    Note: The minimum value achieved for a given evaluation metric (RMSE, MAE, and CRPS) is highlighted in bold.
\end{tablenotes}
\end{threeparttable}
  \label{tab:eu_performance_summary}
\end{table}

\begin{table}[!h]
  \centering
  \caption{Forecasting performance for Switzerland}
   \begin{threeparttable}
  \begin{tabular}{l cccccc}
\hline\hline
 & \multicolumn{3}{c}{$h=1$} & \multicolumn{3}{c}{$h=3$} \bigstrut[t]\\
\cmidrule(lr){2-4} \cmidrule(lr){5-7}
\textbf{Model} & \textbf{RMSE} & \textbf{MAE} & \textbf{CRPS} & \textbf{RMSE} & \textbf{MAE} & \textbf{CRPS} \bigstrut[b]\\
\hline
\multicolumn{7}{l}{\textit{Baseline Models}} \bigstrut[t]\\
Naive last & 0.286 & 0.217 & 0.159 & 0.561 & 0.421 & 0.310 \\
Rolling mean & 0.956 & 0.748 & 0.543 & 1.047 & 0.819 & 0.597 \\
PNC & 0.956 & 0.806 & 0.552 & 1.047 & 0.884 & 0.630 \\
ARIMA(1,1,0) & 0.285 & 0.216 & 0.158 & 0.558 & 0.418 & 0.308 \bigstrut[b]\\
\hline
\multicolumn{7}{l}{\textit{BVAR Models}} \bigstrut[t]\\
BVAR (diffuse) & 0.349 & 0.270 & 0.193 & 0.614 & 0.463 & 0.338 \\
BVAR (Minnesota) & 0.345 & 0.267 & 0.191 & 0.609 & 0.459 & 0.335 \\
BVAR (NW) & 0.344 & 0.266 & 0.191 & 0.612 & 0.463 & 0.337 \bigstrut[b]\\
\hline
\multicolumn{7}{l}{\textit{Multivariate \& ML Models}} \bigstrut[t]\\
DFM & \textbf{0.284} & \textbf{0.215} & \textbf{0.158} & \textbf{0.556} & \textbf{0.414} & \textbf{0.304} \\
DRF & 0.412 & 0.302 & 0.228 & 0.622 & 0.465 & 0.339 \bigstrut\\\hline\hline

 & \multicolumn{3}{c}{$h=6$} & \multicolumn{3}{c}{$h=12$} \bigstrut[t]\\
\cmidrule(lr){2-4} \cmidrule(lr){5-7}
\textbf{Model} & \textbf{RMSE} & \textbf{MAE} & \textbf{CRPS} & \textbf{RMSE} & \textbf{MAE} & \textbf{CRPS} \bigstrut[b]\\
\hline
\multicolumn{7}{l}{\textit{Baseline Models}} \bigstrut[t]\\
Naive last & 0.836 & 0.629 & 0.462 & 1.254 & 0.982 & 0.703 \\
Rolling mean & 1.152 & 0.907 & 0.659 & 1.254 & 0.997 & 0.728 \\
PNC & 1.152 & 0.952 & 0.720 & 1.254 & 1.038 & 0.806 \\
ARIMA(1,1,0) & 0.830 & 0.622 & 0.457 & 1.254 & 0.980 & 0.703 \bigstrut[b]\\
\hline
\multicolumn{7}{l}{\textit{BVAR Models}} \bigstrut[t]\\
BVAR (diffuse) & 0.879 & 0.668 & 0.488 & 1.310 & 1.030 & 0.778 \\
BVAR (Minnesota) & 0.872 & 0.659 & 0.482 & 1.297 & 1.015 & 0.766 \\
BVAR (NW) & 0.877 & 0.667 & 0.488 & 1.308 & 1.028 & 0.781 \bigstrut[b]\\
\hline
\multicolumn{7}{l}{\textit{Multivariate \& ML Models}} \bigstrut[t]\\
DFM & 0.834 & 0.629 & 0.457 & 1.290 & 1.008 & 0.725 \\
DRF & \textbf{0.802} & \textbf{0.588} & \textbf{0.422} & \textbf{0.869} & \textbf{0.619} & \textbf{0.440} \bigstrut\\\hline\hline
\end{tabular}
\vspace*{-0.5cm}
\begin{tablenotes}
    \singlespacing
    \item \leavevmode\kern-\scriptspace\kern-\labelsep 
    Note: The minimum value achieved for a given evaluation metric (RMSE, MAE, and CRPS) is highlighted in bold.
\end{tablenotes}
\end{threeparttable}
  \label{tab:ch_performance_summary}
\end{table}
\FloatBarrier
\subsection{Further calibration results}
\label{app:calibration}

\subsection{United States}

\subsubsection{Short horizons}

\begin{figure}[!h]
  \centering
  \includegraphics[width=\textwidth]{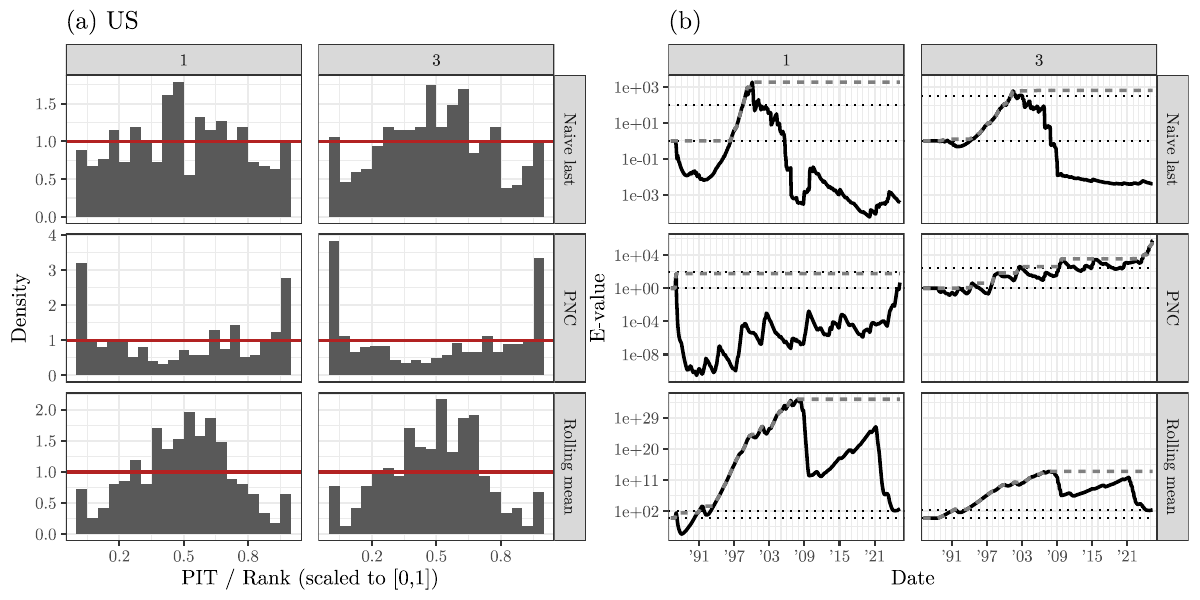}
  \vspace{-2.25em}
  \caption{Calibration diagnostics for Baseline models (Gaussian no-change, Rolling mean, PNC) at short horizons ($h=1,3$) for the United States.  Panel (a) shows PIT/normalized rank histograms. Panel (b) shows cumulative e-values on a log scale; note the different y-scale across rows. The dotted horizontal lines indicate levels of 1 and the rejection thresholds.}
  \label{fig:us_baseline_short}
\end{figure}

\begin{figure}[!h]
  \centering
  \includegraphics[width=\textwidth]{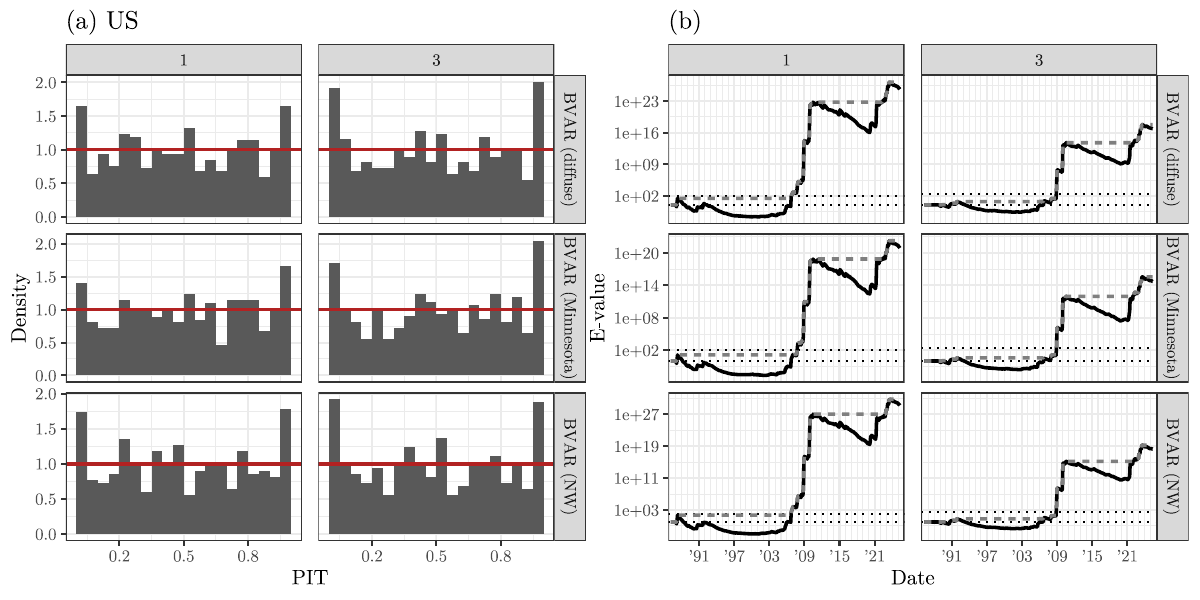}
  \vspace{-2.25em}
  \caption{Calibration diagnostics for BVAR (diffuse, Minnesota, NW) models at short horizons ($h=1,3$) for the United States. Panel (a) shows PIT histograms. Panel (b) shows cumulative e-values on a log scale; note the different y-scale across rows. The dotted horizontal lines indicate levels of 1 and the rejection thresholds.}
  \label{fig:us_bvar_short}
\end{figure}

\clearpage

\subsubsection{Long horizons}

\begin{figure}[!h]
  \centering
  \includegraphics[width=\textwidth]{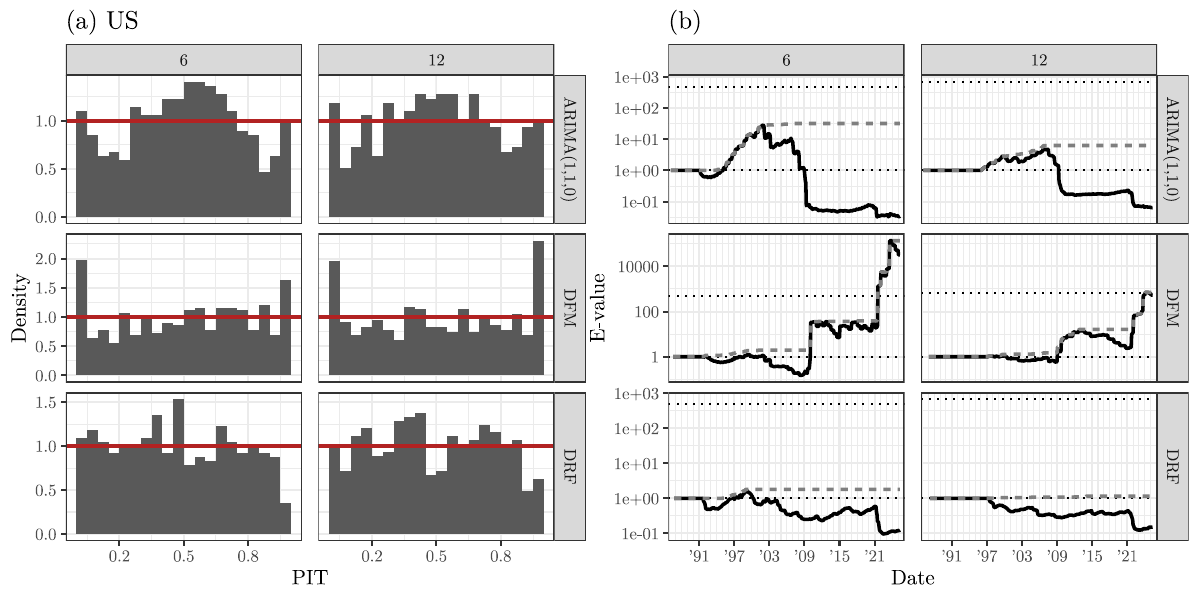}
  \vspace{-2.25em}
  \caption{Calibration diagnostics for ARIMA(1,1,0), DFM and DRF models at long horizons ($h=3,6,12$) for the United States. Panel (a) shows PIT histograms. Panel (b) shows cumulative e-values on a log scale; note the different y-scale across rows. The dotted horizontal lines indicate levels of 1 and the rejection thresholds.}
  \label{fig:us_advanced_long}
\end{figure}

\begin{figure}[!h]
  \centering
  \includegraphics[width=\textwidth]{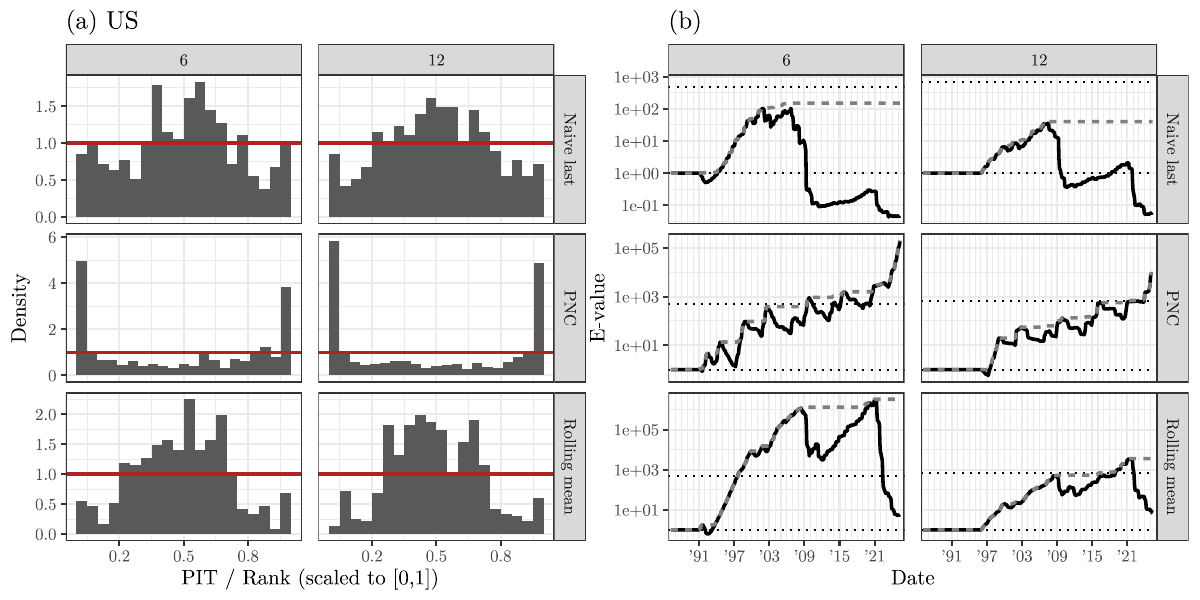}
  \vspace{-2.25em}
  \caption{Calibration diagnostics for Baseline models (Gaussian no-change, Rolling mean, PNC) at long horizons ($h=3,6,12$) for the United States. Panel (a) shows PIT/normalized rank histograms. Panel (b) shows cumulative e-values on a log scale; note the different y-scale across rows. The dotted horizontal lines indicate levels of 1 and the rejection thresholds.}
  \label{fig:us_baseline_long}
\end{figure}

\begin{figure}[!h]
  \centering
  \includegraphics[width=\textwidth]{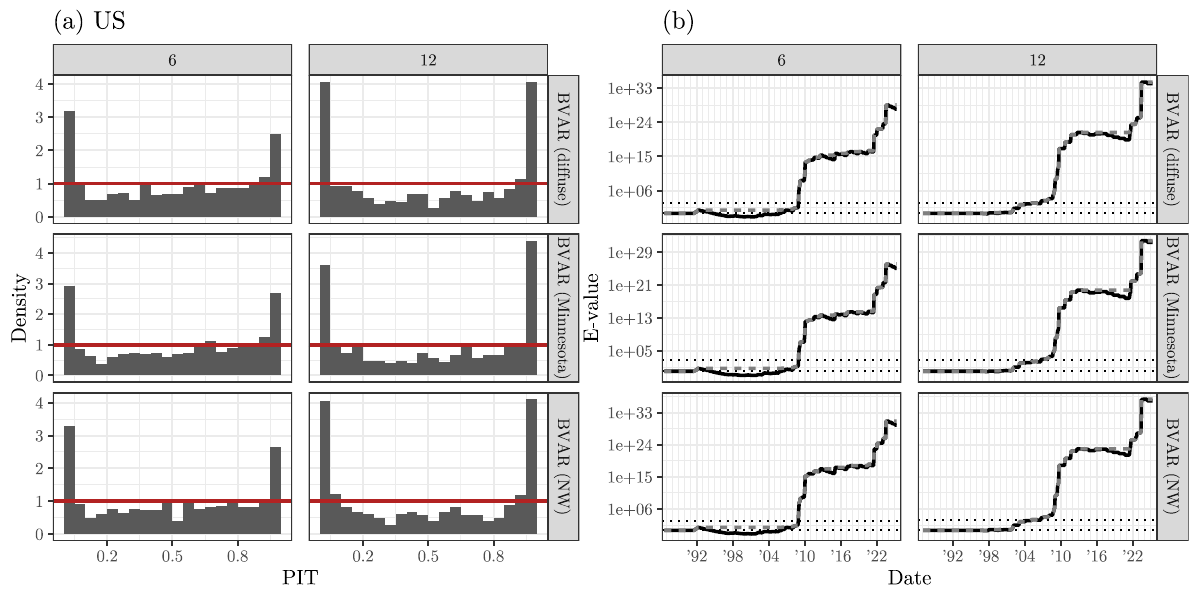}
  \vspace{-2.25em}
  \caption{Calibration diagnostics for BVAR (diffuse, Minnesota, NW) models at long horizons ($h=3,6,12$) for the United States. Panel (a) shows PIT histograms. Panel (b) shows cumulative e-values on a log scale; note the different y-scale across rows. The dotted horizontal lines indicate levels of 1 and the rejection thresholds.}
  \label{fig:us_bvar_long}
\end{figure}

\FloatBarrier

\subsection{Euro Area}

\subsubsection{Short horizons}

\begin{figure}[!h]
  \centering
  \includegraphics[width=\textwidth]{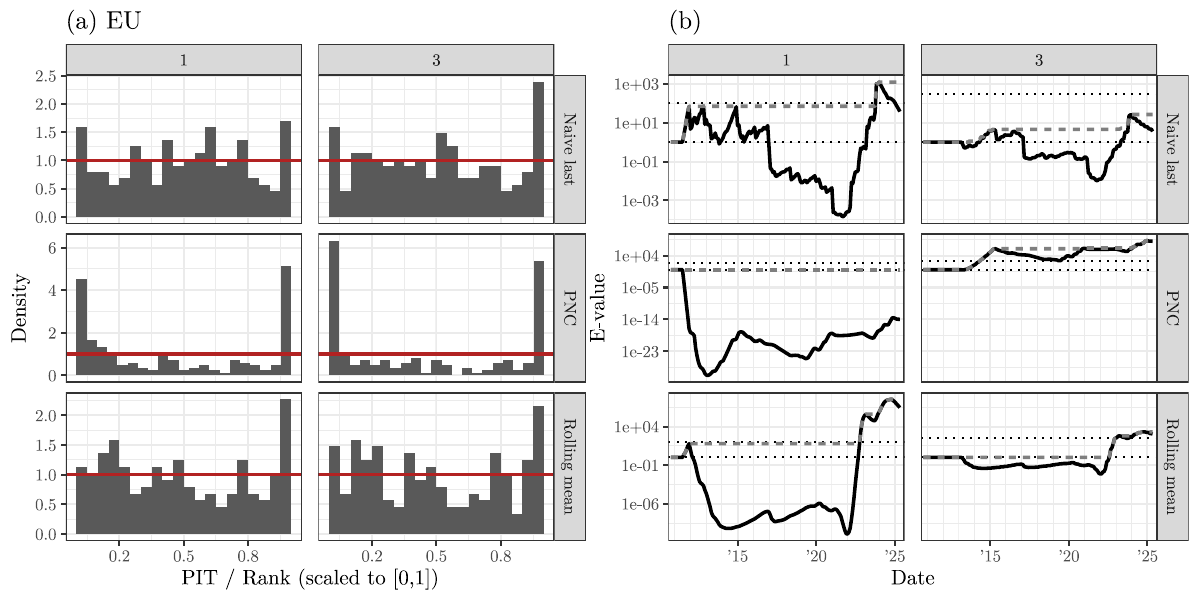}
  \vspace{-2.25em}
  \caption{Calibration diagnostics for Baseline models (Gaussian no-change, Rolling mean, PNC) at short horizons ($h=1,3$) for the Euro Area. Panel (a) shows PIT/normalized rank histograms. Panel (b) shows cumulative e-values on a log scale; note the different y-scale across rows. The dotted horizontal lines indicate levels of 1 and the rejection thresholds.}
  \label{fig:eu_baseline_short}
\end{figure}

\begin{figure}[!h]
  \centering
  \includegraphics[width=\textwidth]{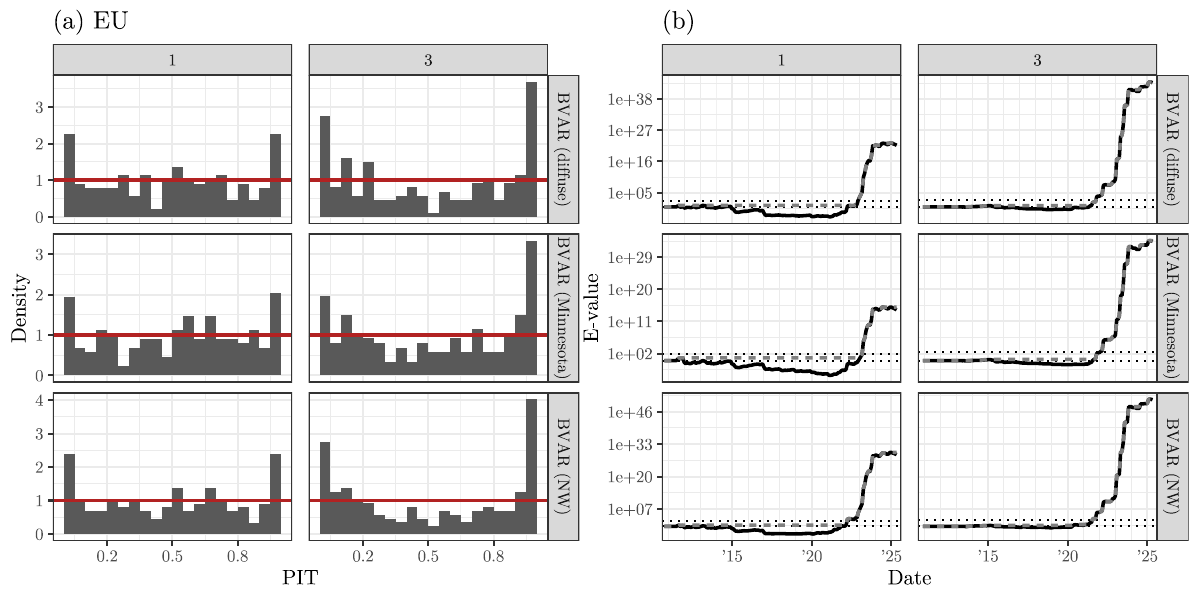}
  \vspace{-2.25em}
  \caption{Calibration diagnostics for BVAR (diffuse, Minnesota, NW) models at short horizons ($h=1,3$) for the Euro Area. Panel (a) shows PIT histograms. Panel (b) shows cumulative e-values on a log scale; note the different y-scale across rows. The dotted horizontal lines indicate levels of 1 and the rejection thresholds.}
  \label{fig:eu_bvar_short}
\end{figure}

\clearpage

\subsubsection{Long horizons}
\begin{figure}[!h]
  \centering
  \includegraphics[width=\textwidth]{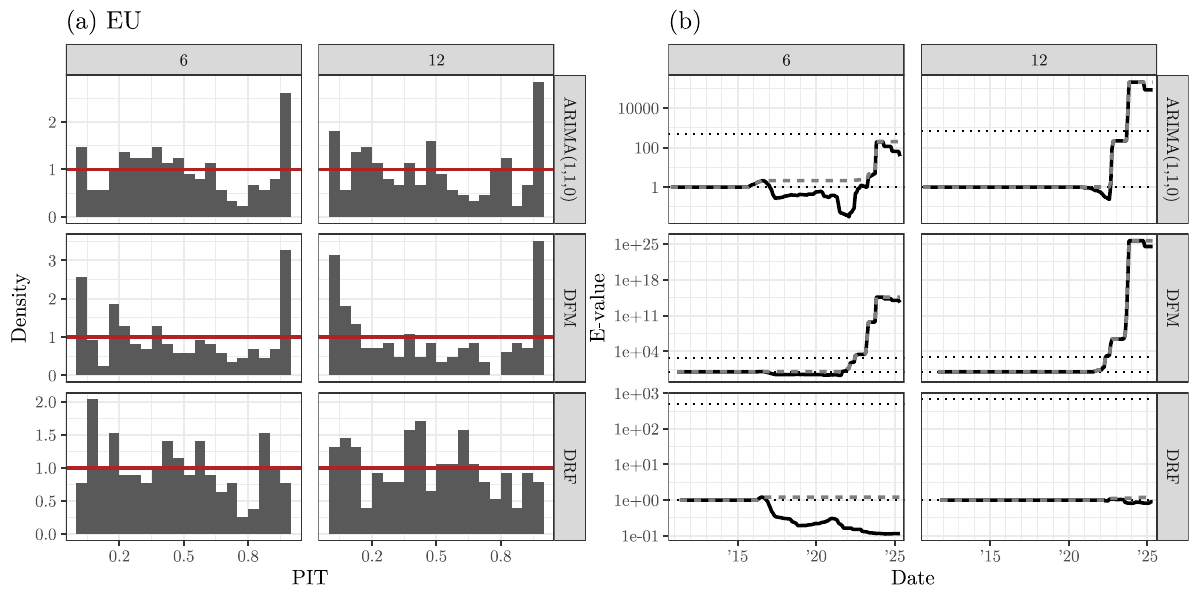}
  \vspace{-2.25em}
  \caption{Calibration diagnostics for ARIMA(1,1,0), DFM and DRF models at long horizons ($h=3,6,12$) for the Euro Area. Panel (a) shows PIT histograms. Panel (b) shows cumulative e-values on a log scale; note the different y-scale across rows. The dotted horizontal lines indicate levels of 1 and the rejection thresholds.}
  \label{fig:eu_advanced_long}
\end{figure}

\begin{figure}[!h]
  \centering
  \includegraphics[width=\textwidth]{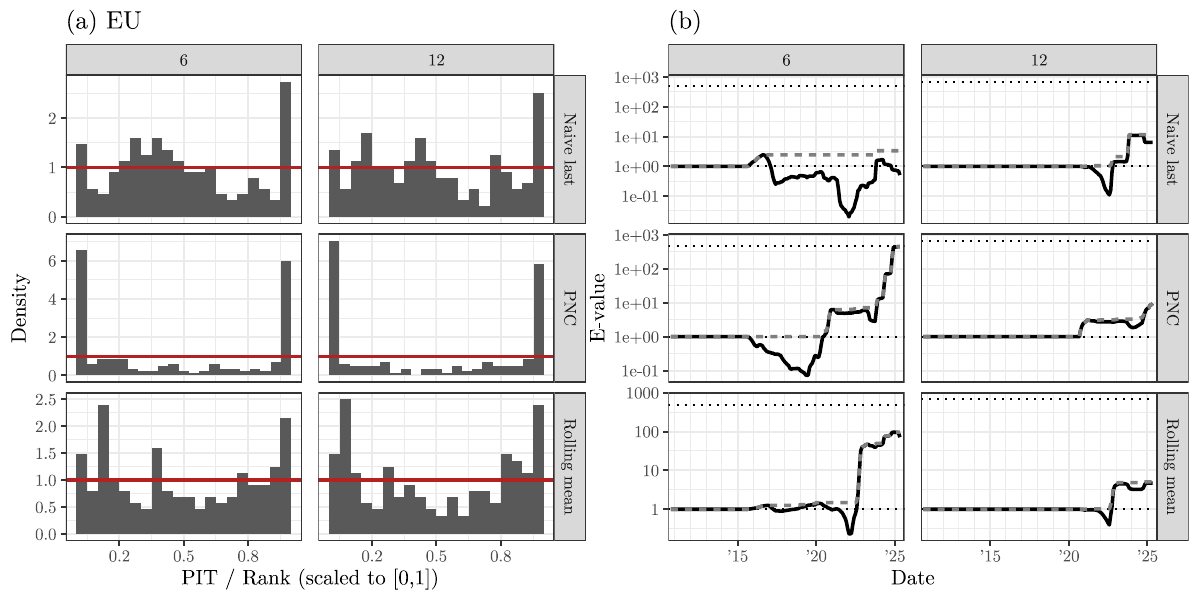}
  \vspace{-2.25em}
  \caption{Calibration diagnostics for Baseline models (Gaussian no-change, Rolling mean, PNC) at long horizons ($h=3,6,12$) for the Euro Area.  Panel (a) shows PIT/normalized rank histograms. Panel (b) shows cumulative e-values on a log scale; note the different y-scale across rows. The dotted horizontal lines indicate levels of 1 and the rejection thresholds.}
  \label{fig:eu_baseline_long}
\end{figure}

\begin{figure}[!h]
  \centering
  \includegraphics[width=\textwidth]{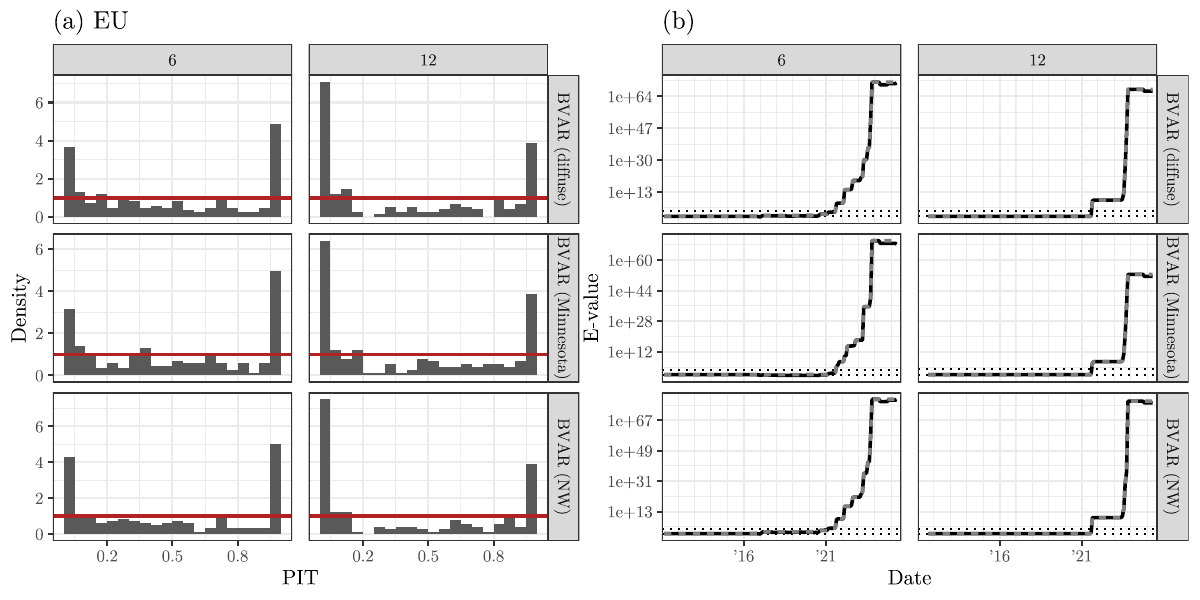}
  \vspace{-2.25em}
  \caption{Calibration diagnostics for BVAR (diffuse, Minnesota, NW) models at long horizons ($h=3,6,12$) for the Euro Area. Panel (a) shows PIT histograms. Panel (b) shows cumulative e-values on a log scale; note the different y-scale across rows. The dotted horizontal lines indicate levels of 1 and the rejection thresholds.}
  \label{fig:eu_bvar_long}
\end{figure}

\clearpage

\begin{table}[!h]
  \centering
  \caption{Calibration diagnostics for the United States}
  \begin{threeparttable}
  \begin{tabular}{l cccccc}
\hline\hline
 & \multicolumn{3}{c}{$h=1$} & \multicolumn{3}{c}{$h=3$} \bigstrut[t]\\
\cmidrule(lr){2-4} \cmidrule(lr){5-7}
\textbf{Model} & \makecell{\textbf{KS Test} \\ \textbf{($\mathbf{p}$)}} & \makecell{\textbf{Edge} \\ \textbf{freq. (\%)}} & \textbf{max $\mathbf{e}$} & \makecell{\textbf{KS Test} \\ \textbf{($\mathbf{p}$)}} & \makecell{\textbf{Edge} \\ \textbf{freq. (\%)}} & \textbf{max $\mathbf{e}$} \bigstrut[b]\\
\hline
\multicolumn{7}{l}{\textit{Baseline Models}} \bigstrut[t]\\
ARIMA(1,1,0) & 0.260 & 0.0 & $1.82\times 10^{4}$ & 0.012 & 0.0 & $7.77\times 10^{1}$ \\
Naive last & 0.158 & 0.0 & $1.87\times 10^{3}$ & 0.005 & 0.0 & $6.48\times 10^{2}$ \\
PNC & -- & 28.5 & $5.68\times 10^{1}$ & -- & 34.2 & $5.44\times 10^{5}$ \\
Rolling mean & 0.000 & 0.0 & $2.28\times 10^{34}$ & 0.000 & 0.0 & $3.50\times 10^{13}$ \bigstrut[b]\\
\hline
\multicolumn{7}{l}{\textit{BVAR Models}} \bigstrut[t]\\
BVAR (Minnesota) & 0.381 & 0.0 & $1.91\times 10^{22}$ & 0.066 & 0.0 & $3.87\times 10^{15}$ \\
BVAR (NW) & 0.307 & 0.0 & $5.66\times 10^{30}$ & 0.072 & 0.0 & $1.71\times 10^{19}$ \\
BVAR (diffuse) & 0.395 & 0.0 & $1.61\times 10^{27}$ & 0.088 & 0.0 & $6.19\times 10^{17}$ \bigstrut[b]\\
\hline
\multicolumn{7}{l}{\textit{Multivariate \& ML Models}} \bigstrut[t]\\
DFM & 0.097 & 0.0 & $9.70\times 10^{4}$ & 0.432 & 0.0 & $3.72\times 10^{5}$ \\
DRF & 0.000 & 1.7 & $4.70\times 10^{14}$ & 0.016 & 1.7 & $2.49\times 10^{1}$ \bigstrut\\
\hline\hline

 & \multicolumn{3}{c}{$h=6$} & \multicolumn{3}{c}{$h=12$} \bigstrut[t]\\
\cmidrule(lr){2-4} \cmidrule(lr){5-7}
\textbf{Model} & \makecell{\textbf{KS Test} \\ \textbf{($\mathbf{p}$)}} & \makecell{\textbf{Edge} \\ \textbf{freq. (\%)}} & \textbf{max $\mathbf{e}$} & \makecell{\textbf{KS Test} \\ \textbf{($\mathbf{p}$)}} & \makecell{\textbf{Edge} \\ \textbf{freq. (\%)}} & \textbf{max $\mathbf{e}$} \bigstrut[b]\\
\hline
\multicolumn{7}{l}{\textit{Baseline Models}} \bigstrut[t]\\
ARIMA(1,1,0) & 0.034 & 0.0 & $3.20\times 10^{1}$ & 0.242 & 0.0 & $6.31\times 10^{0}$ \\
Naive last & 0.001 & 0.0 & $1.52\times 10^{2}$ & 0.003 & 0.0 & $3.99\times 10^{1}$ \\
PNC & -- & 41.8 & $1.99\times 10^{5}$ & -- & 51.0 & $8.79\times 10^{3}$ \\
Rolling mean & 0.000 & 0.0 & $3.21\times 10^{6}$ & 0.000 & 0.0 & $3.53\times 10^{3}$ \bigstrut[b]\\
\hline
\multicolumn{7}{l}{\textit{BVAR Models}} \bigstrut[t]\\
BVAR (Minnesota) & 0.000 & 0.0 & $1.43\times 10^{26}$ & 0.000 & 0.2 & $6.55\times 10^{31}$ \\
BVAR (NW) & 0.000 & 0.0 & $4.01\times 10^{30}$ & 0.000 & 0.4 & $5.17\times 10^{36}$ \\
BVAR (diffuse) & 0.000 & 0.0 & $3.35\times 10^{28}$ & 0.000 & 0.2 & $3.42\times 10^{34}$ \bigstrut[b]\\
\hline
\multicolumn{7}{l}{\textit{Multivariate \& ML Models}} \bigstrut[t]\\
DFM & 0.139 & 0.0 & $1.36\times 10^{5}$ & 0.020 & 0.0 & $7.07\times 10^{2}$ \\
DRF & 0.059 & 1.5 & $1.82\times 10^{0}$ & 0.183 & 1.7 & $1.15\times 10^{0}$ \bigstrut\\
\hline\hline
\end{tabular}
\vspace*{-0.5cm}
\begin{tablenotes}
  \singlespacing
  \item[] Note: The p-values are from a two-sided one-sample KS test for uniformity of the PITs. Edge frequency denotes the share of PIT/rank values at the boundaries during the evaluation period. Max e denotes the maximum value attained by the e-value process over the evaluation period.
\end{tablenotes}
  \end{threeparttable}
  \label{tab:us_calibration_summary}
\end{table}

\begin{table}[!h]
  \centering
  \caption{Calibration diagnostics for the Euro Area}
  \begin{threeparttable}
  \begin{tabular}{l cccccc}
\hline\hline
 & \multicolumn{3}{c}{$h=1$} & \multicolumn{3}{c}{$h=3$} \bigstrut[t]\\
\cmidrule(lr){2-4} \cmidrule(lr){5-7}
\textbf{Model} & \makecell{\textbf{KS Test} \\ \textbf{($\mathbf{p}$)}} & \makecell{\textbf{Edge} \\ \textbf{freq. (\%)}} & \textbf{max $\mathbf{e}$} & \makecell{\textbf{KS Test} \\ \textbf{($\mathbf{p}$)}} & \makecell{\textbf{Edge} \\ \textbf{freq. (\%)}} & \textbf{max $\mathbf{e}$} \bigstrut[b]\\
\hline
\multicolumn{7}{l}{\textit{Baseline Models}} \bigstrut[t]\\
ARIMA(1,1,0) & 0.517 & 0.0 & $2.00\times 10^{2}$ & 0.386 & 0.0 & $1.10\times 10^{2}$ \\
Naive last & 0.657 & 0.0 & $1.21\times 10^{3}$ & 0.147 & 0.0 & $2.63\times 10^{1}$ \\
PNC & -- & 46.0 & $1.00\times 10^{0}$ & -- & 55.7 & $1.50\times 10^{8}$ \\
Rolling mean & 0.147 & 0.0 & $3.37\times 10^{7}$ & 0.117 & 0.0 & $2.04\times 10^{3}$ \bigstrut[b]\\
\hline
\multicolumn{7}{l}{\textit{BVAR Models}} \bigstrut[t]\\
BVAR (Minnesota) & 0.205 & 0.0 & $1.05\times 10^{15}$ & 0.000 & 0.0 & $3.94\times 10^{33}$ \\
BVAR (NW) & 0.161 & 0.0 & $5.07\times 10^{29}$ & 0.000 & 0.0 & $1.04\times 10^{51}$ \\
BVAR (diffuse) & 0.225 & 0.0 & $3.46\times 10^{22}$ & 0.000 & 0.0 & $7.35\times 10^{43}$ \bigstrut[b]\\
\hline
\multicolumn{7}{l}{\textit{Multivariate \& ML Models}} \bigstrut[t]\\
DFM & 0.815 & 0.0 & $1.90\times 10^{7}$ & 0.100 & 0.0 & $8.91\times 10^{13}$ \\
DRF & 0.004 & 6.3 & $1.13\times 10^{2}$ & 0.057 & 6.4 & $6.03\times 10^{0}$ \bigstrut\\
\hline\hline

 & \multicolumn{3}{c}{$h=6$} & \multicolumn{3}{c}{$h=12$} \bigstrut[t]\\
\cmidrule(lr){2-4} \cmidrule(lr){5-7}
\textbf{Model} & \makecell{\textbf{KS Test} \\ \textbf{($\mathbf{p}$)}} & \makecell{\textbf{Edge} \\ \textbf{freq. (\%)}} & \textbf{max $\mathbf{e}$} & \makecell{\textbf{KS Test} \\ \textbf{($\mathbf{p}$)}} & \makecell{\textbf{Edge} \\ \textbf{freq. (\%)}} & \textbf{max $\mathbf{e}$} \bigstrut[b]\\
\hline
\multicolumn{7}{l}{\textit{Baseline Models}} \bigstrut[t]\\
ARIMA(1,1,0) & 0.073 & 0.0 & $2.00\times 10^{2}$ & 0.046 & 0.0 & $2.05\times 10^{5}$ \\
Naive last & 0.040 & 0.0 & $3.34\times 10^{0}$ & 0.120 & 0.0 & $1.17\times 10^{1}$ \\
PNC & -- & 59.7 & $4.51\times 10^{2}$ & -- & 61.4 & $9.54\times 10^{0}$ \\
Rolling mean & 0.111 & 0.0 & $9.61\times 10^{1}$ & 0.010 & 0.0 & $5.06\times 10^{0}$ \bigstrut[b]\\
\hline
\multicolumn{7}{l}{\textit{BVAR Models}} \bigstrut[t]\\
BVAR (Minnesota) & 0.000 & 0.6 & $1.39\times 10^{70}$ & 0.000 & 9.1 & $3.33\times 10^{52}$ \\
BVAR (NW) & 0.000 & 1.8 & $3.01\times 10^{79}$ & 0.000 & 9.7 & $2.29\times 10^{78}$ \\
BVAR (diffuse) & 0.000 & 1.8 & $3.81\times 10^{71}$ & 0.000 & 9.1 & $6.04\times 10^{67}$ \bigstrut[b]\\
\hline
\multicolumn{7}{l}{\textit{Multivariate \& ML Models}} \bigstrut[t]\\
DFM & 0.009 & 0.0 & $4.20\times 10^{14}$ & 0.000 & 0.0 & $4.28\times 10^{25}$ \\
DRF & 0.065 & 7.1 & $1.20\times 10^{0}$ & 0.154 & 6.7 & $1.28\times 10^{0}$ \bigstrut\\
\hline\hline
\end{tabular}
\vspace*{-0.5cm}
\begin{tablenotes}
  \singlespacing
  \item[] Note: The p-values are from a two-sided one-sample KS test for uniformity of the PITs. Edge frequency denotes the share of PIT/rank values at the boundaries during the evaluation period. Max e denotes the maximum value attained by the e-value process over the evaluation period.
\end{tablenotes}
  \end{threeparttable}
  \label{tab:eu_calibration_summary}
\end{table}

\FloatBarrier

\section{Evaluation of probabilistic inflation forecasts for Switzerland}
\label{chap:results_ch}

\subsection{Results for the Baseline models}
\label{sec:baseline_results_ch}
\begin{figure}[!t]
  \centering
  \includegraphics[width=1\textwidth]{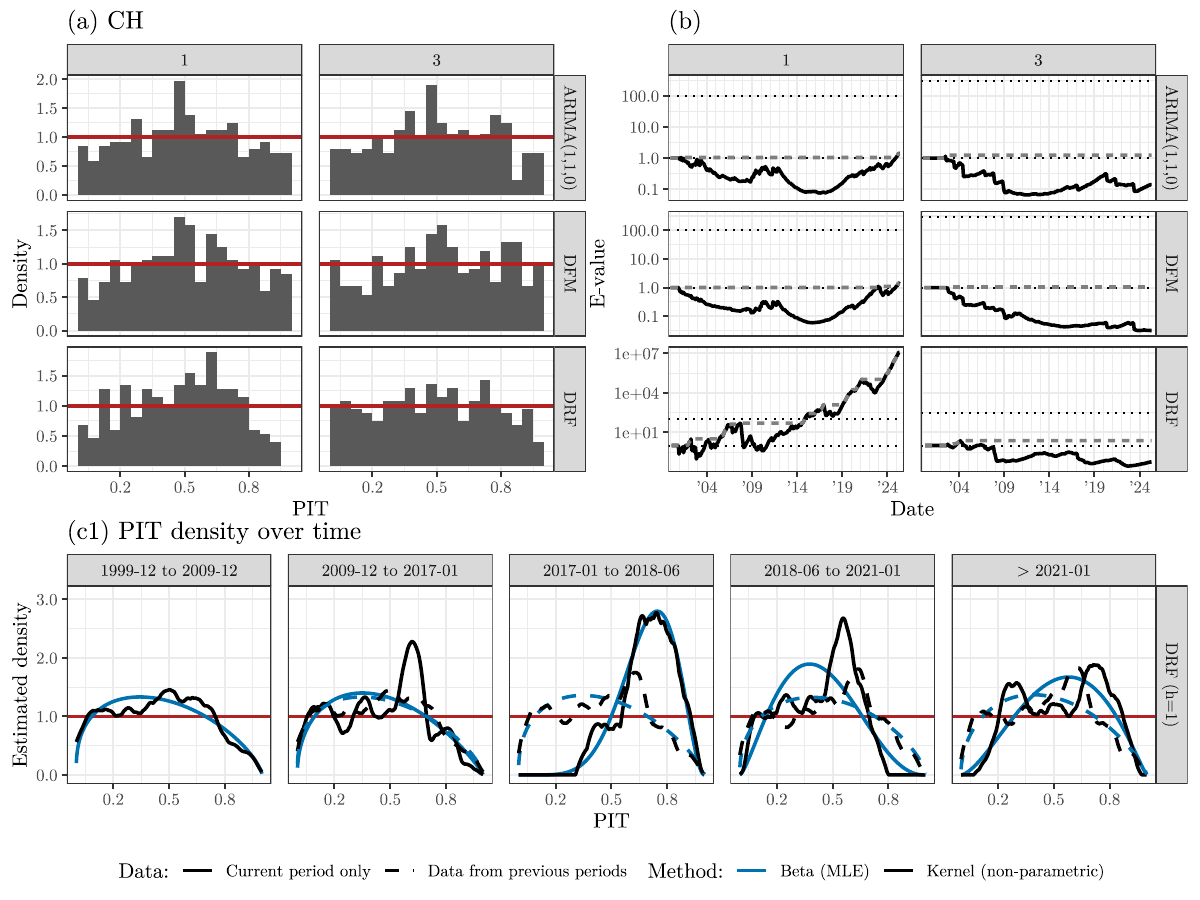}
  \vspace{-2.25em}
  \caption{Calibration diagnostics for ARIMA(1,1,0), DFM and DRF models at short horizons ($h=1,3$) for Switzerland. Panel (a) shows PIT histograms. Panel (b) shows cumulative e-values on a log scale; note the different y-scale across rows. The dotted horizontal lines indicate levels of 1 and the rejection thresholds. Panel (c) shows the evolution of the density estimates of the PIT.}
  \label{fig:ch_advanced_short}
\end{figure}

\begin{figure}[!t]
  \centering
  \includegraphics[width=\textwidth]{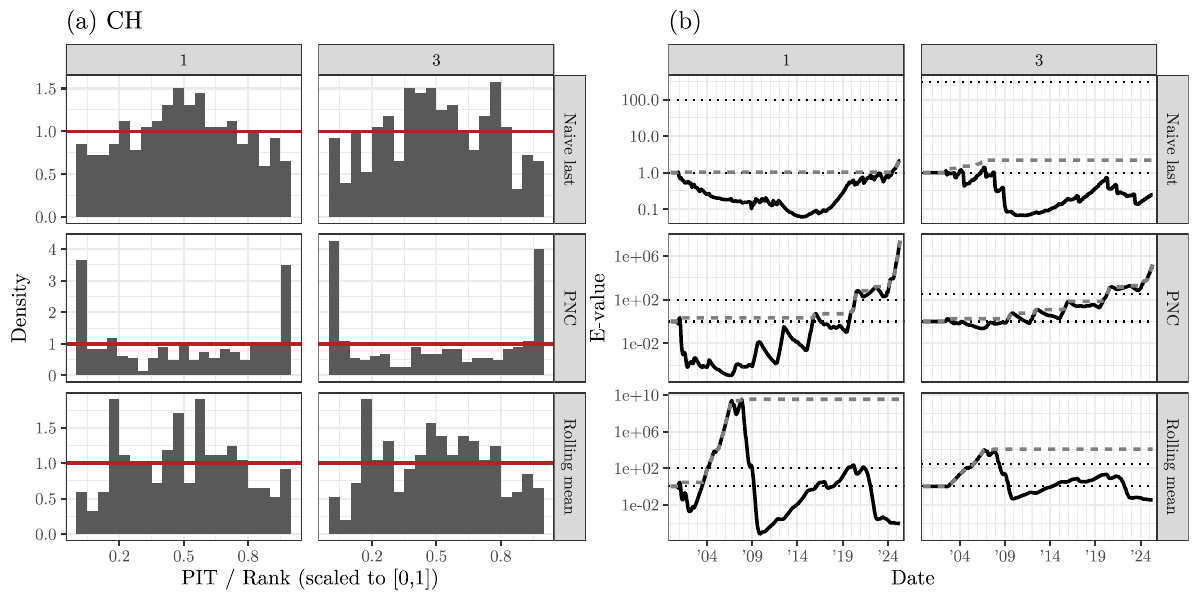}
  \vspace{-2.25em}
  \caption{Calibration diagnostics for Baseline models (Gaussian no-change, Rolling mean, PNC) at short horizons ($h=1,3$) for Switzerland. Panel (a) shows PIT/normalized rank histograms. Panel (b) shows cumulative e-values on a log scale; note the different y-scale across rows. The dotted horizontal lines indicate levels of 1 and the rejection thresholds.}
  \label{fig:ch_baseline_short}
\end{figure}

\begin{figure}[!t]
  \centering
  \includegraphics[width=\textwidth]{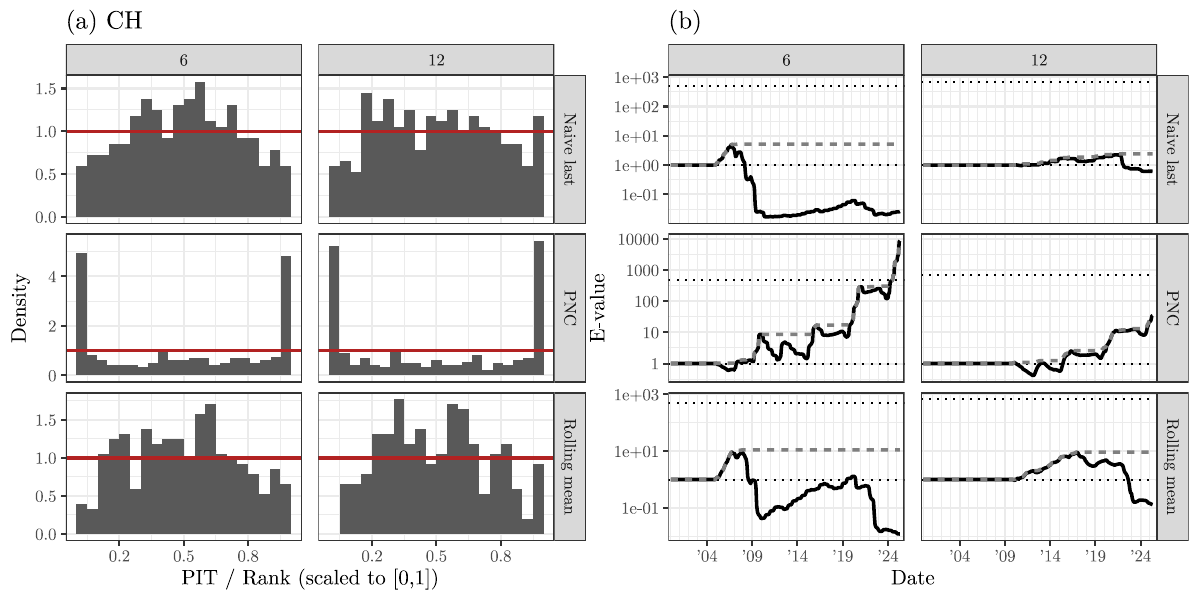}
  \vspace{-2.25em}
  \caption{Calibration diagnostics for Baseline models (Gaussian no-change, Rolling mean, PNC) at long horizons ($h=3,6,12$) for Switzerland. Panel (a) shows PIT/normalized rank histograms. Panel (b) shows cumulative e-values on a log scale; note the different y-scale across rows. The dotted horizontal lines indicate levels of 1 and the rejection thresholds.}
  \label{fig:ch_baseline_long}
\end{figure}

For Switzerland, the ARIMA(1,1,0) and Gaussian no-change forecasts appear calibrated across all horizons. The PIT histograms for these methods are only mildly inverse U-shaped. Their cumulative e-value processes do not approach our rejection threshold; see Figures~\ref{fig:ch_advanced_short}, \ref{fig:ch_baseline_short}, \ref{fig:ch_advanced_long}, and~\ref{fig:ch_baseline_long}. The traditional one-sample KS test also does not reject uniformity of the PITs, which supports this assessment (Table~\ref{tab:ch_calibration_summary}).

By contrast, the rolling mean forecast is clearly miscalibrated (Figures~\ref{fig:ch_baseline_short} and~\ref{fig:ch_baseline_long}). Although its PIT histograms do not look much worse than those of the better calibrated models, the one-sample KS test rejects calibration at all horizons (Table~\ref{tab:ch_calibration_summary}). The sequential analysis confirms this result and adds temporal detail. For the one-month horizon ($h=1$), the cumulative e-value process crosses our rejection threshold and reaches values of order $10^{9}$ (Figure~\ref{fig:ch_baseline_short}). For $h=3$, it provides strong evidence against calibration, growing to values of order $10^{4}$. Figure~\ref{fig:ch_baseline_short} also shows that most of this evidence accumulates between 2002 and 2008. The e-values then fall rapidly until 2010 and increase slowly until 2020 before declining again. For longer horizons ($h \geq 6$), the cumulative e-values no longer signal a problem, which is consistent with the expected loss of power for multi-step forecasts (Figure~\ref{fig:ch_baseline_long}).

Finally, the PNC method does not produce calibrated forecasts. Its rank histograms place a large share of ranks in the first and last bins, rising from 34\% for $h=1$ to over 50\% for $h=12$. Yet the cumulative e-values respond only slowly to this severe miscalibration (Figures~\ref{fig:ch_baseline_short} and~\ref{fig:ch_baseline_long}, Table~\ref{tab:ch_calibration_summary}).

\subsection{Results for the BVAR models}
\label{sec:bvar_results_ch}

\begin{figure}[!h]
  \centering
  \includegraphics[width=\textwidth]{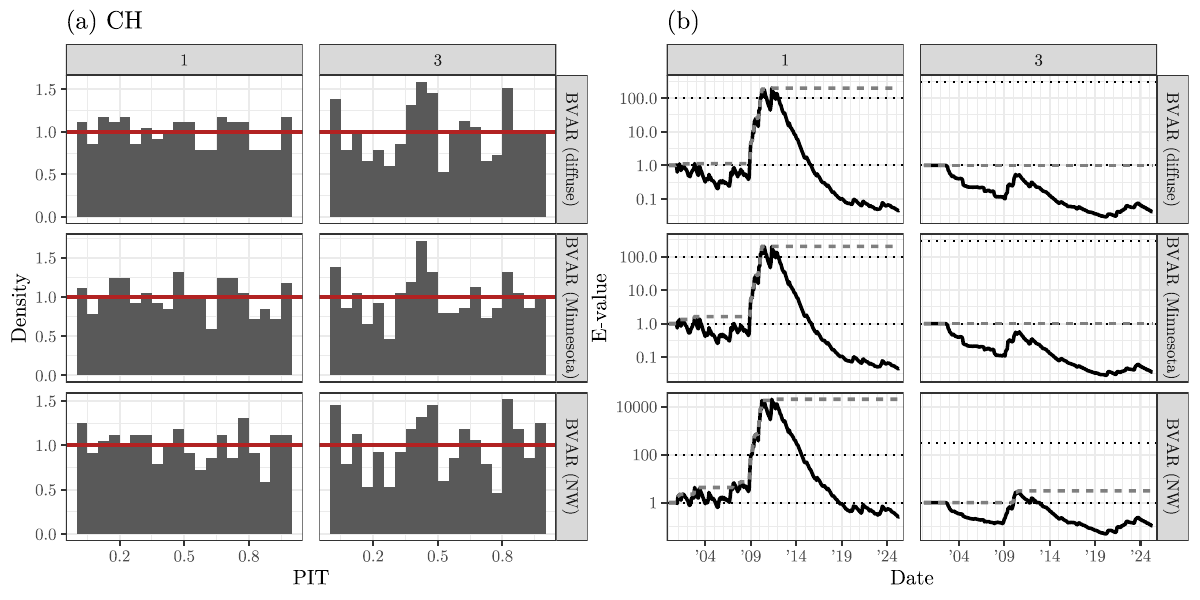}
  \vspace{-2.25em}
  \caption{Calibration diagnostics for BVAR (diffuse, Minnesota, NW) models at short horizons ($h=1,3$) for Switzerland. Panel (a) shows PIT histograms. Panel (b) shows cumulative e-values on a log scale; note the different y-scale across rows. The dotted horizontal lines indicate levels of 1 and the rejection thresholds.}
  \label{fig:ch_bvar_short}
\end{figure}

The sequential analysis for Switzerland provides a nuanced, horizon-dependent view of this miscalibration. For the one-month horizon ($h=1$), Figure~\ref{fig:ch_bvar_short} shows that the sequential method uncovers miscalibration that the static KS test misses (Table~\ref{tab:ch_calibration_summary}). Although the full-sample PIT histogram appears fairly uniform and the one-sample KS test does not reject calibration, the test supermartingale reveals a significant problem, reaching levels of order $10^{2}$ and $10^{4}$ during the Great Financial Crisis (GFC).

\begin{figure}[!t]
  \centering
  \includegraphics[width=\textwidth]{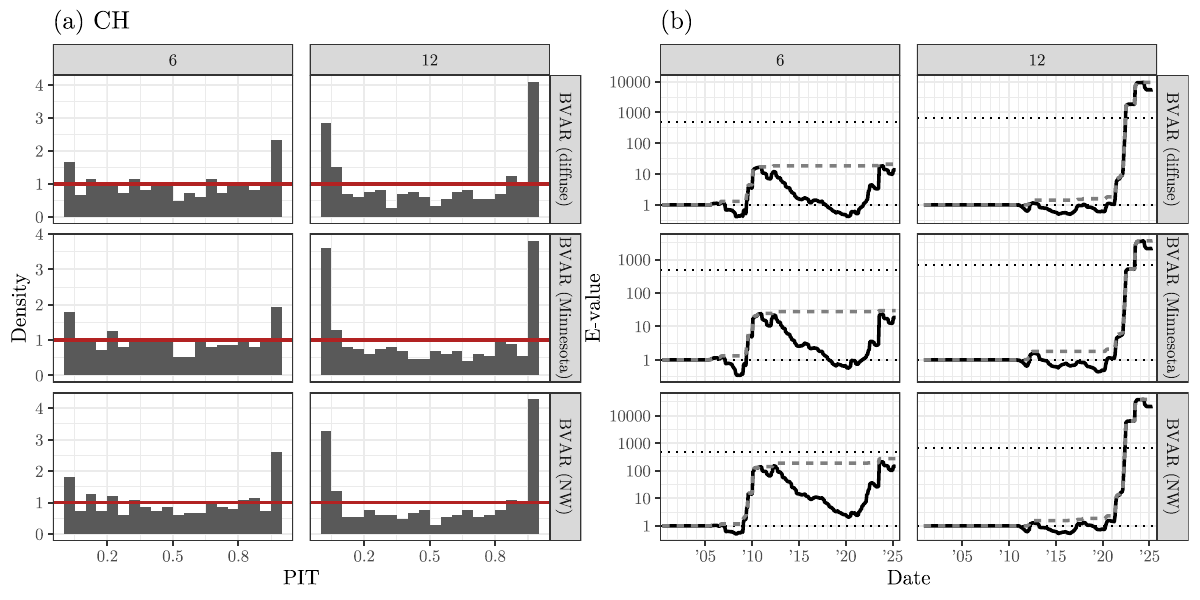}
  \vspace{-2.25em}
  \caption{Calibration diagnostics for BVAR (diffuse, Minnesota, NW) models at long horizons ($h=3,6,12$) for Switzerland. Panel (a) shows PIT histograms. Panel (b) shows cumulative e-values on a log scale; note the different y-scale across rows. The dotted horizontal lines indicate levels of 1 and the rejection thresholds.}
  \label{fig:ch_bvar_long}
\end{figure}

For the three-month horizon, the BVAR forecasts appear to be better calibrated, since both the one-sample KS test and the cumulative e-values find no evidence against calibration. In fact, this is the only horizon-region combination where the BVARs are clearly not miscalibrated. However, the miscalibration re-emerges at longer horizons (Figure~\ref{fig:ch_bvar_long}). For $h=6$ and especially for $h=12$, the PIT histograms show pronounced bunching of values at 0 and 1. This underlying miscalibration is severe enough to be detected by the sequential test, despite its power loss at longer horizons. For the 12-month forecasts, the cumulative e-values indicate serious misspecification, reaching levels of order $10^{3}$ and $10^{5}$ during the COVID-19 period. This finding is consistent with the static KS test, which rejects calibration (with p-values effectively equal to zero) for all priors for the longest horizon (Table~\ref{tab:ch_calibration_summary}).

\subsection{Results for the DFM}
\label{sec:dfm_results_ch}
\begin{figure}[!t]
  \centering
  \includegraphics[width=\textwidth]{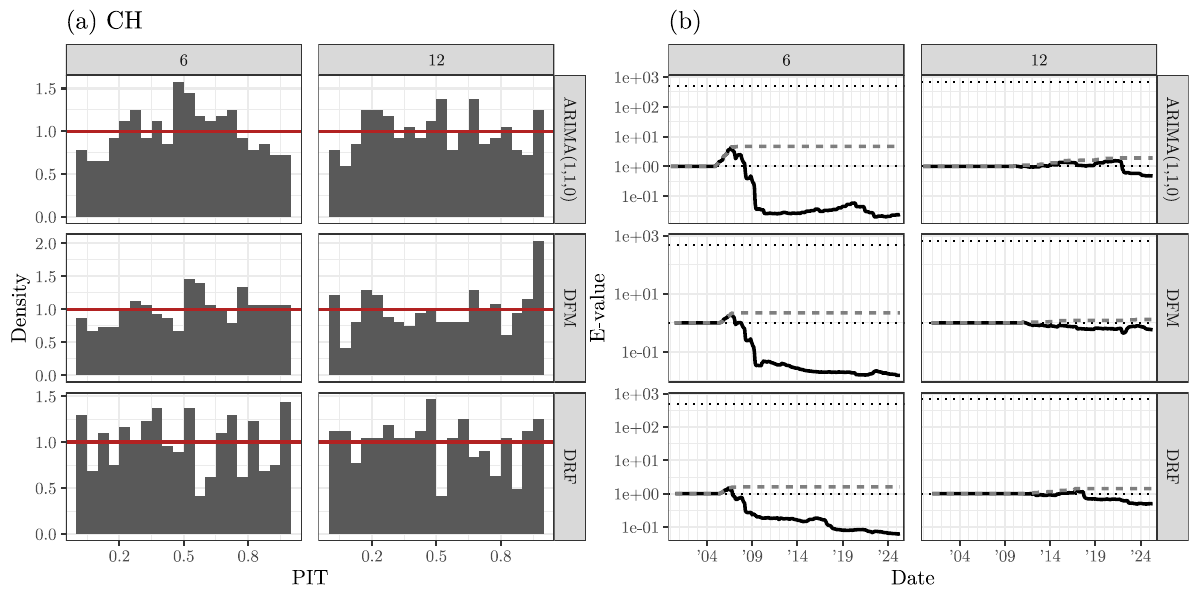}
  \vspace{-2.25em}
  \caption{Calibration diagnostics for ARIMA(1,1,0), DFM and DRF models at long horizons ($h=3,6,12$) for Switzerland. Panel (a) shows PIT histograms. Panel (b) shows cumulative e-values on a log scale; note the different y-scale across rows. The dotted horizontal lines indicate levels of 1 and the rejection thresholds.}
  \label{fig:ch_advanced_long}
\end{figure}

For Switzerland, the DFM maintained calibration for extended periods. Notably, the cumulative e-values never indicate evidence against calibration at any horizon, a result that mirrors the calibration of the ARIMA model (Figures~\ref{fig:ch_advanced_short} and~\ref{fig:ch_advanced_long}). The static KS test gives p-values between $0.035$ (for $h=3$) and $0.158$ (for $h=12$), suggesting some evidence against calibration. However, this evidence is never decisive (Table~\ref{tab:ch_calibration_summary}).

\subsection{Results for the DRF}
\label{sec:drf_results_ch}
For Switzerland, for instance, the DRF's calibration is most problematic for the shortest horizon. For one-month-ahead forecasts ($h=1$), the process decisively rejects calibration, reaching a level of order $10^7$. The corresponding PIT histogram in panel~(a) of Figure~\ref{fig:ch_advanced_short} shows an inverse U-shape, indicating that the miscalibration comes from an overdispersion in the forecast distributions. However, the process in panel~(b) does not show a linear, steady increase. To investigate this, we again plot the estimated densities of the alternative hypotheses over time in panel~(c1). We choose the subperiods according to the major developments of the process. From 2000 to 2009, the process shows some movement, but does not cross $100$. The estimated beta density from only this period shows that the DRF forecasts were indeed overdispersed. This is again the case from 2010 to 2017, where the process accumulates enough evidence to reject calibration on the grounds of overdispersion. However, we observe a short decrease from 2017 to 2018. The distribution of the alternative during this period shows that this decrease is not because of better calibration. In fact, the forecasts are biased during this time. The process does not increase initially because it only has power if the data is consistent with the current alternative distribution. The middle plot in panel~(c1) shows that the current alternative (dotted line) tests mainly for overdispersion. For the rest of the evaluation period, the estimated alternative adapts and the process grows further.

In contrast, for all $h \geq 3$, the e-values provide no evidence against calibration (Figure~\ref{fig:ch_advanced_long}). The static KS test largely supports this horizon-dependent finding. It rejects uniformity of the PITs for $h=1$ with a p-value of essentially zero and fails to reject it for $h \geq 3$ with p-values greater than $0.164$ (Table~\ref{tab:ch_calibration_summary}).

\begin{table}[!h]
  \centering
  \caption{Calibration diagnostics for Switzerland}
  \begin{threeparttable}
  \begin{tabular}{l cccccc}
\hline\hline
 & \multicolumn{3}{c}{$h=1$} & \multicolumn{3}{c}{$h=3$} \bigstrut[t]\\
\cmidrule(lr){2-4} \cmidrule(lr){5-7}
\textbf{Model} & \makecell{\textbf{KS Test} \\ \textbf{($\mathbf{p}$)}} & \makecell{\textbf{Edge} \\ \textbf{freq. (\%)}} & \textbf{max $\mathbf{e}$} & \makecell{\textbf{KS Test} \\ \textbf{($\mathbf{p}$)}} & \makecell{\textbf{Edge} \\ \textbf{freq. (\%)}} & \textbf{max $\mathbf{e}$} \bigstrut[b]\\
\hline
\multicolumn{7}{l}{\textit{Baseline Models}} \bigstrut[t]\\
ARIMA(1,1,0) & 0.109 & 0.0 & $1.42\times 10^{0}$ & 0.068 & 0.0 & $1.24\times 10^{0}$ \\
Naive last & 0.198 & 0.0 & $2.13\times 10^{0}$ & 0.093 & 0.0 & $2.19\times 10^{0}$ \\
PNC & -- & 34.1 & $2.44\times 10^{7}$ & -- & 39.3 & $1.59\times 10^{5}$ \\
Rolling mean & 0.021 & 0.0 & $3.32\times 10^{9}$ & 0.042 & 0.0 & $1.21\times 10^{4}$ \bigstrut[b]\\
\hline
\multicolumn{7}{l}{\textit{BVAR Models}} \bigstrut[t]\\
BVAR (Minnesota) & 0.871 & 0.0 & $2.03\times 10^{2}$ & 0.633 & 0.0 & $1.00\times 10^{0}$ \\
BVAR (NW) & 0.966 & 0.0 & $2.00\times 10^{4}$ & 0.580 & 0.0 & $3.06\times 10^{0}$ \\
BVAR (diffuse) & 0.864 & 0.0 & $2.00\times 10^{2}$ & 0.369 & 0.0 & $1.00\times 10^{0}$ \bigstrut[b]\\
\hline
\multicolumn{7}{l}{\textit{Multivariate \& ML Models}} \bigstrut[t]\\
DFM & 0.055 & 0.0 & $1.44\times 10^{0}$ & 0.035 & 0.0 & $1.03\times 10^{0}$ \\
DRF & 0.000 & 2.3 & $1.21\times 10^{7}$ & 0.164 & 2.3 & $2.37\times 10^{0}$ \bigstrut\\
\hline\hline

 & \multicolumn{3}{c}{$h=6$} & \multicolumn{3}{c}{$h=12$} \bigstrut[t]\\
\cmidrule(lr){2-4} \cmidrule(lr){5-7}
\textbf{Model} & \makecell{\textbf{KS Test} \\ \textbf{($\mathbf{p}$)}} & \makecell{\textbf{Edge} \\ \textbf{freq. (\%)}} & \textbf{max $\mathbf{e}$} & \makecell{\textbf{KS Test} \\ \textbf{($\mathbf{p}$)}} & \makecell{\textbf{Edge} \\ \textbf{freq. (\%)}} & \textbf{max $\mathbf{e}$} \bigstrut[b]\\
\hline
\multicolumn{7}{l}{\textit{Baseline Models}} \bigstrut[t]\\
ARIMA(1,1,0) & 0.186 & 0.0 & $4.70\times 10^{0}$ & 0.536 & 0.0 & $1.91\times 10^{0}$ \\
Naive last & 0.098 & 0.0 & $5.20\times 10^{0}$ & 0.158 & 0.0 & $2.43\times 10^{0}$ \\
PNC & -- & 46.6 & $8.79\times 10^{3}$ & -- & 50.8 & $3.50\times 10^{1}$ \\
Rolling mean & 0.035 & 0.0 & $1.11\times 10^{1}$ & 0.004 & 0.0 & $9.30\times 10^{0}$ \bigstrut[b]\\
\hline
\multicolumn{7}{l}{\textit{BVAR Models}} \bigstrut[t]\\
BVAR (Minnesota) & 0.220 & 0.0 & $2.97\times 10^{1}$ & 0.000 & 0.0 & $3.64\times 10^{3}$ \\
BVAR (NW) & 0.017 & 0.0 & $2.74\times 10^{2}$ & 0.000 & 0.0 & $3.93\times 10^{4}$ \\
BVAR (diffuse) & 0.085 & 0.0 & $2.09\times 10^{1}$ & 0.000 & 0.0 & $9.60\times 10^{3}$ \bigstrut[b]\\
\hline
\multicolumn{7}{l}{\textit{Multivariate \& ML Models}} \bigstrut[t]\\
DFM & 0.058 & 0.0 & $2.20\times 10^{0}$ & 0.158 & 0.0 & $1.31\times 10^{0}$ \\
DRF & 0.549 & 2.3 & $1.61\times 10^{0}$ & 0.319 & 2.4 & $1.42\times 10^{0}$ \bigstrut\\
\hline\hline
\end{tabular}
\vspace*{-0.5cm}
\begin{tablenotes}
  \singlespacing
  \item[] Note: The p-values are from a two-sided one-sample KS test for uniformity of the PITs. Edge frequency denotes the share of PIT/rank values at the boundaries during the evaluation period. Max e denotes the maximum value attained by the e-value process over the evaluation period.
\end{tablenotes}
  \end{threeparttable}
  \label{tab:ch_calibration_summary}
\end{table}

\FloatBarrier

\section{Details on the Forecasting Models} \label{app:models}

This section describes the models used to generate the probabilistic forecasts for the empirical application.

\subsection{Baseline forecasts}
\label{sec:baseline_forecasts}
We begin by introducing a set of simple benchmark forecasts that serve as reference points for the evaluation of the more advanced models considered subsequently. We implement three baseline methods: a non-parametric, probabilistic no-change forecast (PNC), a Gaussian random walk forecast (``naive last''), and a simple rolling-mean forecast.

\subsubsection{Probabilistic no-change forecast}
\label{ssec:pnc}
The PNC forecast provides a straightforward non-parametric benchmark \citep{gneiting_predicting_2010}. It assumes that the recent empirical distribution of inflation represents the best predictor for its future distribution. This approach makes no assumption about distributional form and can capture skewness, multimodality, or heavy tails. 

We construct the forecast using a rolling window of the most recent $L=20$ observations, following \citet{gneiting_predicting_2010}. The predictive CDF for an observation $y$ is
\[
\hat{F}_{\mathrm{PNC},t}(y)
=\frac{1}{L}\sum_{j=1}^{L} \mathbf{1}_{\{\pi_{t-j} \le y\}}.
\]
The window length balances adaptability and stability. A shorter window reacts faster to changes but is noisier. A longer window provides smoother yet slower adjustments.

\subsubsection{Gaussian no-change forecast}
\label{ssec:gaussian_no_change}
The Gaussian no-change forecast assumes that the inflation follows a random walk,
\[
\pi_{t+1} = \pi_t + \varepsilon_{t+1},
\]
where $\varepsilon_{t+1}$ is a zero-mean innovation. Given this assumption, the best forecast for $\pi_{t+h}$, made at time $t$, is simply $\pi_t$.

To extend this random walk model to a probabilistic forecast, we assume Gaussian forecast errors. We do not impose the theoretical $h$-step variance ($h$ times the one-step variance) following \citet{hyndman_forecasting_2021}. Instead, we estimate the forecast uncertainty empirically. Specifically, for each horizon $h$, we issue the forecast
\[
Y_{t+h} \mid \mathcal{F}_t \sim \mathcal{N}(\pi_t, \hat{\sigma}_h^2),
\]
where $\hat{\sigma}_h^2$ is estimated from the mean squared $h$-step residuals obtained from all previous forecasts. This expanding-window estimation uses the historical distribution of forecast errors for each horizon. It captures the true realized uncertainty of past forecasts and is therefore robust to violations of the strict random walk assumption. This empirical calibration of $\hat{\sigma}_h^2$ ensures that forecast intervals reflect observed volatility rather than theoretical scaling. Note that this construction of the forecast uncertainty mirrors the approach used by the ECB and the Fed for their published density projections.

\subsubsection{Rolling mean forecast}
\label{ssec:rolling_mean}
The rolling mean forecast smooths recent observations to reduce short-term noise and emphasise the local trend \citep{hyndman_forecasting_2021}. It assumes that inflation is locally stationary within a short window. The point forecast for horizon $h$ at time $t$ is the arithmetic mean of the most recent $L=20$ values:
\[
\hat{\pi}_{t+h \mid t} = \bar{\pi}_t = \frac{1}{L}\sum_{j=0}^{L-1} \pi_{t-j}.
\]
We again assume Gaussian forecast errors and define the predictive distribution as
\[
Y_{t+h} \mid \mathcal{F}_t \sim \mathcal{N}(\hat{\pi}_{t+h \mid t}, \hat{\sigma}_h^2).
\]
We estimate the variance $\hat{\sigma}_h^2$ separately for each horizon $h$ based on the historical performance of $h$-step rolling mean forecasts. The variance incorporates two sources of uncertainty. First, the residual uncertainty reflects the inherent unpredictability of inflation for horizon $h$. Second, the parameter uncertainty arises because $\bar{\pi}_t$ estimates the local mean from a finite sample. Standard forecasting principles require that both components contribute to the total predictive variance. Following \citet{hyndman_forecasting_2021}, we therefore apply a scaling factor of $\sqrt{1+1/L}$ to the empirical residual standard deviation. This adjustment ensures that the predictive intervals reflect both the inherent randomness of inflation and the estimation uncertainty of the rolling mean.

\subsection{ARIMA forecast}
\label{ssec:arima}
Further, we implement an autoregressive integrated moving average (ARIMA$(p,d,q)$) model.

\subsubsection{Model selection}
Using the inflation series observed up to time $t$, we determine the order of integration $d$ via the augmented Dickey–Fuller (ADF) test, which rejects the null hypothesis of non-stationarity for the inflation series for all three countries after first differencing. The lag orders $(p,q)$ are selected using the Bayesian information criterion (BIC), with a search over $p \le 5$ and $q \le 5$.

In our application, the ARIMA(1,1,0) model is selected more often than any other specification. It chooses this model in roughly 81\% of cases for Switzerland, 48\% for the EA, and 44\% for the US, where ARIMA(0,1,2) appears more often. Restricting the US sample to post-2000 data raises the ARIMA(1,1,0) share to 67\%. For consistency across datasets, we adopt ARIMA(1,1,0) as our baseline specification, leading to the following model specification
\[
\Delta \pi_t = \alpha_0 + \phi_1 \Delta \pi_{t-1} + \varepsilon_t,
\]
where $\Delta \pi_t = \pi_t - \pi_{t-1}$ and $\varepsilon_t$ is a white-noise error.

\subsubsection{Forecasting workflow and probabilistic forecasts}
We use an expanding window to re-estimate the ARIMA(1,1,0) model at each cutoff date~$t$. At each $t$, we update the estimated parameters $\alpha_0$ and $\phi_1$ using all available data up to that month. We then generate iterated forecasts for horizons $h \in \{1,3,6,12\}$.

By assuming Gaussian forecast errors, the predictive distribution for the inflation is given by
\[
Y_{t+h} \mid \mathcal{F}_t \sim \mathcal{N}(\hat{\pi}_{t+h \mid t}, \hat{\sigma}_h^2),
\]
where $\hat{\pi}_{t+h \mid t}$ denotes the mean forecast for horizon $h$. This parametrization yields a probabilistic forecast that integrates model-based uncertainty and allows direct comparison with the other approaches.

\subsection{Bayesian vector autoregression}
\label{sec:bvar}
Univariate models provide useful benchmarks but assume that the target evolves independently of other indicators. In macroeconomics, variables like inflation, monetary aggregates, and interest rates co-move and influence each other over time. Vector autoregressions (VARs) capture these dynamic interdependencies. They model each variable as a function of its own lags and the lags of other variables in the system \citep{sims_macroeconomics_1980}. 

A standard VAR suffers from the ``curse of dimensionality'': the number of parameters grows quadratically in the number of variables and linearly in the lag order. With many variables and limited history, we face unstable estimation and weak forecast performance \citep{koop_bayesian_2010}. Bayesian VARs address this problem by introducing priors that shrink coefficients toward parsimonious reference dynamics, for example, white noise or a random walk.

\subsubsection{The VAR(p) model and Bayesian shrinkage}
Let $\mathbf{y}_t \in \mathbb{R}^M$ collect $M$ stationary series at time $t$. A VAR($p$) with intercept is
\[
\mathbf{y}_t = \boldsymbol{\alpha}_0 + \sum_{j=1}^{p} \mathbf{A}_j \mathbf{y}_{t-j} + \boldsymbol{\varepsilon}_t, 
\qquad \boldsymbol{\varepsilon}_t \sim \mathcal{N}(\mathbf{0}, \boldsymbol{\Sigma}),
\]
where $\boldsymbol{\alpha}_0 \in \mathbb{R}^M$ and $\mathbf{A}_j \in \mathbb{R}^{M \times M}$. Stack $\mathbf{x}_t = (1, \mathbf{y}_{t-1}^\prime, \dots, \mathbf{y}_{t-p}^\prime)$ and define $\mathbf{X} = (x_1^\prime,\dots,x_T^\prime) \in \mathbb{R}^{T \times K}$ with $K = 1+Mp$. Let $\mathbf{A} = (\boldsymbol{\alpha}_0, \mathbf{A}_1, \dots, \mathbf{A}_p)^\prime \in \mathbb{R}^{K \times M}$ and stack $\mathbf{Y}=(\mathbf{y}_1,\dots,\mathbf{y}_T)^\prime \in \mathbb{R}^{T \times M}$. Then, we obtain:
\[
\mathbf{Y} = \mathbf{X}\mathbf{A} + \mathbf{E}, \qquad \mathrm{vec}(\mathbf{E}) \sim \mathcal{N}\!\left(\mathbf{0},\, \mathbf{I}_T \otimes \boldsymbol{\Sigma}\right).
\]
Bayesian inference treats $\mathbf{A}$ and possibly $\boldsymbol{\Sigma}$ as random. It combines a prior with the Gaussian likelihood and yields a posterior for the parameters and a predictive distribution that averages over parameter uncertainty.

\subsubsection{Model specification and forecasting process}
\label{ssec:bvarspecs}
We estimate BVARs on the transformed, stationary data defined in Section~\ref{sec:data}. For each country, we use the full predictor set and follow an expanding-window scheme. At each monthly forecast origin~$t$, we re-estimate the model on all data up to~$t$ and produce iterated $h$-step forecasts for $h \in \{1,3,6,12\}$.

We generate multi-step forecasts by iterating one-step predictions on the stationary system. Since we estimate the BVAR on differences (or otherwise transformed series), the iterated predictions yield forecasts of future changes. We recover the level forecast for inflation by cumulating the predicted changes and adding them to the last observed level at the forecast origin. To quantify the uncertainty of this multi-step forecast, we compute a cumulative variance. This cumulative variance equals the sum of the individual forecast error variances from step~1 through step~$h$. Under the standard assumption that forecast errors at each step are independent, this procedure aligns the predictive variance with the cumulative point forecast \citep{marcellino_comparison_2006}.

\subsubsection{Lag Length Selection}
We select the lag order $p$ by a data-driven procedure. At regular intervals between the 40\% and 90\% sample quantiles, we fit VARs with $p\in\{1,\dots,12\}$ and compute the BIC. The BIC selects $p=1$ at all grid points for Switzerland and the US. For the EA, it mostly selects $p=1$ at most grid points but returns higher orders at early sample points, because the number of parameters exceeds the number of observations. To ensure comparability across regions, we set $p=1$ for all BVARs.

\subsubsection{Prior specifications}
\label{ssec:priors}
We consider three priors: The diffuse prior serves as a no-shrinkage baseline. The Minnesota prior imposes equation-wise Gaussian shrinkage under a fixed diagonal error covariance. The Normal-inverse-Wishart prior (often called Normal-Wishart in the precision parametrization) is the natural conjugate prior that treats both coefficients and the covariance matrix as random. All three yield closed-form posteriors and allow fast recursive re-estimation in our rolling scheme \citep{koop_bayesian_2010}.

\subsubsection{Diffuse prior}
The diffuse (non-informative) prior imposes no shrinkage. The posterior means coincide with OLS equation-by-equation estimates, and we use these as a baseline to measure the gains from shrinkage.

Under this prior, the posterior mean of the coefficient matrix equals the OLS estimator, $\mathbf{A}_{\text{post}}=(\mathbf{X}'\mathbf{X})^{-1}\mathbf{X}'\mathbf{Y}$. We estimate the posterior covariance of the innovations as
\[
\boldsymbol{\Sigma}_{\text{post}} = \frac{(\mathbf{Y}-\mathbf{X}\mathbf{A}_{\text{post}})'(\mathbf{Y}-\mathbf{X}\mathbf{A}_{\text{post}})}{T-K}.
\]
The covariance of the coefficient vector follows as $\mathrm{Var}(\mathrm{vec}(\mathbf{A}))=\boldsymbol{\Sigma}_{\text{post}}\otimes(\mathbf{X}'\mathbf{X})^{-1}$. 
For any regressor vector $\mathbf{x}_{t+1}$, the one-step predictive distribution of $\mathbf{y}_{t+1}$ is multivariate Normal:
\[
\mathbf{y}_{t+1}\mid\mathcal{F}_t \sim 
\mathcal{N}\!\left(\mathbf{x}_{t+1}'\mathbf{A}_{\text{post}},\,
(\mathbf{I}_M\otimes \mathbf{x}_{t+1}')\,(\boldsymbol{\Sigma}_{\text{post}}\otimes(\mathbf{X}'\mathbf{X})^{-1})\,(\mathbf{I}_M\otimes \mathbf{x}_{t+1})+\boldsymbol{\Sigma}_{\text{post}}\right).
\]
We obtain iterated forecasts for horizons $h>1$ by updating the regressor $\mathbf{x}_{t+h}$ sequentially using the most recent predicted values and applying the same one-step density at each horizon. The predictive variance for the cumulative $h$-step forecast is the sum of the variances of the individual step-ahead predictions.

\subsubsection{Minnesota prior}
The Minnesota prior \citep{doan_forecasting_1984,litterman_forecasting_1986} is a widely used shrinkage prior in macroeconomic forecasting. It balances simplicity and flexibility by shrinking each equation in the VAR toward a univariate autoregressive process. This structure reflects the belief that, in the absence of evidence to the contrary, each variable is best predicted by its own past values.

The prior achieves strong shrinkage while retaining analytical tractability because it assumes a fixed, known, and diagonal error covariance matrix $\boldsymbol{\Sigma}$. This assumption implies that the error terms across equations are contemporaneously uncorrelated, and it allows us to estimate each equation separately. While this simplification neglects potential cross-equation correlations, it greatly reduces computational burden and makes the prior well-suited for large systems. We estimate the fixed covariance as $\hat{\boldsymbol{\Sigma}} = \mathrm{diag}(s_1^2, \dots, s_M^2)$, where $s_i^2$ is the OLS residual variance from equation $i$.

We rewrite the VAR in stacked regression form. Let $\mathbf{y} = \mathrm{vec}(\mathbf{Y})$ and $\boldsymbol{\varepsilon} = \mathrm{vec}(\mathbf{E})$. Then
\[
\mathbf{y} = (\mathbf{I}_M \otimes \mathbf{X}) \boldsymbol{\alpha} + \boldsymbol{\varepsilon}, 
\qquad 
\boldsymbol{\varepsilon} \sim \mathcal{N}(\mathbf{0}, \hat{\boldsymbol{\Sigma}} \otimes \mathbf{I}_T),
\]
where $\boldsymbol{\alpha} = \mathrm{vec}(\mathbf{A})$ stacks all coefficients. The Minnesota prior specifies a Normal distribution for $\boldsymbol{\alpha}$:
\[
\boldsymbol{\alpha} \sim \mathcal{N}(\underline{\boldsymbol{\alpha}}_{\mathrm{Mn}}, \underline{\mathbf{V}}_{\mathrm{Mn}}),
\qquad 
\underline{\boldsymbol{\alpha}}_{\mathrm{Mn}} = \mathbf{0}.
\]
Since our time series are stationary, centering the prior at zero shrinks coefficients towards the belief that each series follows a white-noise process. The prior covariance matrix $\underline{\mathbf{V}}_{\mathrm{Mn}}$ is diagonal and determines the strength of shrinkage for different coefficients. The Minnesota prior scales the individual prior variances as follows:
\[
\underline{\mathbf{V}}_{\mathrm{Mn}}(i,j,l) =
\begin{cases}
	\dfrac{\lambda_1}{l^{2}} & \text{for the $l^{th}$ own lag of variable $i$,}\\[6pt]
	\dfrac{\lambda_2\, s_i^{2}}{l^{2}\, s_j^{2}} & \text{for the $l^{th}$ lag of variable $j \neq i$ in equation $i$,}\\[6pt]
	\lambda_3\, s_i^{2} & \text{for the intercept in equation $i$,}
\end{cases}
\]
where $\lambda_1$, $\lambda_2$, and $\lambda_3$ are hyperparameters that control overall tightness. The quadratic decay $1/l^2$ reflects the prior belief that distant lags matter less than recent ones. The ratio $s_i^2/s_j^2$ rescales the shrinkage for variables with different volatilities and makes the prior invariant to units of measurement.

The posterior distribution of the coefficients is Normal because both the prior and likelihood are Gaussian. Let $\hat{\boldsymbol{\alpha}}$ denote the OLS estimate. We then obtain the posterior covariance and mean as
\[
\overline{\mathbf{V}}_{\mathrm{Mn}} 
= 
\left(\underline{\mathbf{V}}_{\mathrm{Mn}}^{-1} + \hat{\boldsymbol{\Sigma}}^{-1}\!\otimes\!(\mathbf{X}'\mathbf{X})\right)^{-1},
\qquad
\overline{\boldsymbol{\alpha}}_{\mathrm{Mn}} 
= 
\overline{\mathbf{V}}_{\mathrm{Mn}}\!\left(\underline{\mathbf{V}}_{\mathrm{Mn}}^{-1}\underline{\boldsymbol{\alpha}}_{\mathrm{Mn}} + (\hat{\boldsymbol{\Sigma}}^{-1}\!\otimes\!\mathbf{X}'\mathbf{X})\hat{\boldsymbol{\alpha}}\right).
\]
Given the Gaussian form of both the posterior and the likelihood, the predictive distribution under the Minnesota prior is also Gaussian. For a regressor vector $\mathbf{x}_{t+1}$, the one-step predictive distribution is
\[
\mathbf{y}_{t+1}\mid\mathcal{F}_t \sim 
\mathcal{N}\!\left(\mathbf{x}_{t+1}'\overline{\mathbf{A}}_{\mathrm{Mn}},\,
(\mathbf{I}_M\otimes \mathbf{x}_{t+1}')\,\overline{\mathbf{V}}_{\mathrm{Mn}}\,(\mathbf{I}_M\otimes \mathbf{x}_{t+1})+\hat{\boldsymbol{\Sigma}}\right),
\]
where $\overline{\mathbf{A}}_{\mathrm{Mn}}$ is the posterior mean coefficient matrix and $\hat{\boldsymbol{\Sigma}}$ the fixed diagonal innovation covariance. For forecast horizons $h>1$, we iterate one-step predictions and update $\mathbf{x}_{t+h}$ recursively with the generated forecasts. The predictive densities at all horizons are therefore Normal, with means generated by this iterative scheme. For cumulative $h$-step forecasts, we obtain the predictive variance as the sum of the one-step-ahead variances along the forecast path, under the standard assumption that forecast errors at each step are independent.

\subsubsection{Normal-Wishart prior}
The Normal-Wishart prior, also known as the natural conjugate prior for the VAR, provides a fully Bayesian treatment of all model parameters. The prior jointly models the coefficient matrix $\mathbf{A}$ and the covariance matrix $\boldsymbol{\Sigma}$ \citep{koop_bayesian_2010}. Its main feature is conjugacy: the posterior belongs to the same distributional family as the prior, which yields closed-form posterior distributions. This property makes the prior a powerful and computationally convenient alternative to simulation-based inference methods.

The conjugate structure arises from the Gaussian likelihood of the VAR. Conditional on $\boldsymbol{\Sigma}$, the likelihood of $\mathbf{A}$ is Normal, and the marginal likelihood of $\boldsymbol{\Sigma}$ follows a Wishart distribution in its precision form. Specifically,
\[
\mathbf{A} \mid \boldsymbol{\Sigma}, \mathbf{Y} 
\sim 
\mathcal{N}\!\big(\hat{\mathbf{A}},\, \boldsymbol{\Sigma} \otimes (\mathbf{X}'\mathbf{X})^{-1}\big),
\qquad
\boldsymbol{\Sigma}^{-1} \mid \mathbf{Y} 
\sim 
W(\mathbf{S}^{-1}, T-K-M-1),
\]
where $\hat{\mathbf{A}}$ is the OLS estimate and $\mathbf{S} = (\mathbf{Y}-\mathbf{X}\hat{\mathbf{A}})'(\mathbf{Y}-\mathbf{X}\hat{\mathbf{A}})$ is the residual sum-of-squares matrix. The prior mirrors this structure:
\[
\mathbf{A} \mid \boldsymbol{\Sigma} 
\sim 
\mathcal{N}\!\big(\underline{\mathbf{A}},\, \boldsymbol{\Sigma} \otimes \underline{\mathbf{V}}\big),
\qquad
\boldsymbol{\Sigma}^{-1} 
\sim 
W(\underline{\mathbf{S}}^{-1}, \underline{\nu}),
\]
with hyperparameters $(\underline{\mathbf{A}}, \underline{\mathbf{V}}, \underline{\mathbf{S}}, \underline{\nu})$. The prior mean matrix $\underline{\mathbf{A}}$ and scale matrix $\underline{\mathbf{S}}$ determine the center of the prior beliefs, while $\underline{\mathbf{V}}$ and $\underline{\nu}$ govern prior variance and degrees of freedom. In our application, we set the prior mean close to zero and use relatively non-informative hyperparameters to allow the data to dominate inference.

Combining the prior with the likelihood yields the posterior parameters:
\begin{align*}
	\overline{\mathbf{V}} &= (\underline{\mathbf{V}}^{-1} + \mathbf{X}'\mathbf{X})^{-1}, \quad
	\overline{\mathbf{A}} = \overline{\mathbf{V}}\!\left(\underline{\mathbf{V}}^{-1}\underline{\mathbf{A}} + \mathbf{X}'\mathbf{Y}\right), \quad
	\overline{\nu} = \underline{\nu} + T,\\
	\overline{\mathbf{S}} &= \mathbf{S} + \underline{\mathbf{S}} + \hat{\mathbf{A}}'\mathbf{X}'\mathbf{X}\hat{\mathbf{A}} + \underline{\mathbf{A}}'\underline{\mathbf{V}}^{-1}\underline{\mathbf{A}} - \overline{\mathbf{A}}'(\underline{\mathbf{V}}^{-1} + \mathbf{X}'\mathbf{X})\overline{\mathbf{A}}.
\end{align*}
The posterior distributions are
\[
\mathbf{A} \mid \boldsymbol{\Sigma}, \mathbf{Y} 
\sim 
\mathcal{N}\!\big(\overline{\mathbf{A}},\, \boldsymbol{\Sigma} \otimes \overline{\mathbf{V}}\big),
\qquad
\boldsymbol{\Sigma}^{-1} \mid \mathbf{Y} 
\sim 
W(\overline{\mathbf{S}}^{-1}, \overline{\nu}).
\]

This fully Bayesian framework integrates over both coefficient and covariance uncertainty when it generates forecasts. The one-step predictive distribution for $\mathbf{y}_{t+1}$, conditional on data up to time~$t$, follows a multivariate $t$-distribution with $\overline{\nu}$ degrees of freedom, mean $\mathbf{x}_{t+1}'\overline{\mathbf{A}}$, and covariance
\[
\frac{1}{\overline{\nu}-M-1}\,\overline{\mathbf{S}}\big(1 + \mathbf{x}_{t+1}'\overline{\mathbf{V}}\mathbf{x}_{t+1}\big).
\]
This expression provides an analytical predictive density that fully accounts for parameter uncertainty. For horizons $h>1$, however, no analytical form exists. We obtain forecasts at these horizons either by simulation, drawing future paths from the posterior, or by specifying direct multi-horizon models. To maintain computational tractability in our recursive forecasting design and to ensure comparability across priors, we use the iterated approach based on the posterior mean coefficients $\overline{\mathbf{A}}$ to approximate the predictive density for $h>1$.

Given the closed-form Normal-Wishart posterior, we approximate the one-step predictive $t$-distribution by a Gaussian density with the same mean and covariance. For a regressor vector $\mathbf{x}_{t+1}$, we write
\[
\mathbf{y}_{t+1}\mid\mathcal{F}_t \sim 
\mathcal{N}\!\left(\mathbf{x}_{t+1}'\overline{\mathbf{A}},\, 
(\mathbf{I}_M\otimes \mathbf{x}_{t+1}')\,(\boldsymbol{\Sigma}_{\text{post}}\otimes \overline{\mathbf{V}})\,(\mathbf{I}_M\otimes \mathbf{x}_{t+1})+\boldsymbol{\Sigma}_{\text{post}}\right),
\]
where $\boldsymbol{\Sigma}_{\text{post}} = \overline{\mathbf{S}}/(\overline{\nu}-M-1)$. For the one-step horizon, the exact predictive density is multivariate $t$, but we approximate it by a Gaussian distribution with identical mean and covariance. This Normal approximation preserves the first two moments of the exact multivariate $t$-distribution and simplifies comparison across models. For forecast horizons $h > 1$, we iterate one-step predictions and update the regressors recursively. This iterative scheme generates the predictive mean. For cumulative $h$-step forecasts, we approximate the predictive variance by summing the one-step conditional variances along the forecast path, under the standard assumption that forecast errors at each step are independent.

\subsection{Dynamic factor model}
\label{sec:dfm}
BVARs are useful for forecasting a small set of related variables. They become difficult to estimate when the dataset is large. DFMs address this problem. They summarize the common movements in many time series with only a few unobserved factors \citep{stock_forecasting_2002,stock_dynamic_2016}. Each factor represents a shared source of variation that affects many variables at once, while the remaining part of each series reflects variable-specific shocks. This approach allows us to use a large amount of information without estimating too many parameters. DFMs are a primary tool for short-term forecasting and nowcasting. They can also handle practical data issues, for example, different publication lags for different variables or different data frequencies \citep{clements_nowcasting_2012}.

\subsubsection{The DFM in state-space form}
Following \citet{banbura_maximum_2014}, we express the DFM in a state-space form. This representation allows estimation and forecasting through the Kalman filter. The model has two equations. The measurement equation links the $M$ observed stationary time series, $\mathbf{y}_t \in \mathbb{R}^M$, to the $k$ latent factors, $\mathbf{f}_t \in \mathbb{R}^k$:
\[
\mathbf{y}_t = \boldsymbol{\Lambda}\mathbf{f}_t + \boldsymbol{\varepsilon}_t,
\]
where $\boldsymbol{\Lambda}$ is the $M \times k$ matrix of factor loadings, and $\boldsymbol{\varepsilon}_t$ contains the idiosyncratic components. Each idiosyncratic term follows an autoregressive process,
\[
\boldsymbol{\varepsilon}_t = \mathbf{R}\boldsymbol{\varepsilon}_{t-1} + \boldsymbol{\eta}_t, \qquad \boldsymbol{\eta}_t \sim \mathcal{N}(\mathbf{0}, \mathbf{H}),
\]
which captures variable-specific persistence and volatility.

The state equation describes how the latent factors evolve over time. The factors follow a VAR($p$) process,
\[
\mathbf{f}_t = \mathbf{A}_1\mathbf{f}_{t-1} + \dots + \mathbf{A}_p\mathbf{f}_{t-p} + \mathbf{u}_t, \qquad \mathbf{u}_t \sim \mathcal{N}(\mathbf{0}, \mathbf{Q}),
\]
where $\mathbf{Q}$ is the covariance matrix of the factor innovations. The rank of $\mathbf{Q}$, denoted $q$, determines the number of independent shocks, or dynamic factors, driving the system, with $q \leq k$.

\subsubsection{Model specification and hyperparameter selection}
\label{ssec:dfmspecs}
The specification of a dynamic factor model depends on three parameters: the number of static factors ($k$), the lag order of the factor VAR process ($p$), and the number of dynamic factors ($q$). We estimate these values from the data using formal statistical criteria. 

Specifically, $k$ is determined using the eigenvalue-ratio test of \citet{onatski_determining_2010}, which is robust to cross-sectional dependence in the idiosyncratic components. The number of dynamic factors $q$ is estimated following \citet{bai_determining_2007}. The lag order is set to $p=1$, as higher-order specifications do not affect the estimated factor structure in our data and do not yield additional explanatory gain.

\subsubsection{Estimation and forecasting process}
We estimate the model by maximum likelihood with the expectation-maximization (EM) algorithm. The EM-algorithm alternates between estimating the latent factors and updating the model parameters until convergence \citep{banbura_maximum_2014}. The forecasting process mirrors the BVAR workflow. We estimate the DFM on transformed, stationary data, so its direct output is a forecast of future changes in the inflation rate. We obtain a forecast for the inflation level in two steps.

First, we generate forecasts for inflation changes from the estimated state-space model with the Kalman filter. At each forecast origin~$t$, we re-estimate the model on the expanding data window that includes all observations up to~$t$. We then compute one-step and multi-step forecasts by iterating the prediction equations of the state-space system. This procedure yields a sequence of predicted changes for horizons $h = 1,2,\dots,12$.

Second, we construct the predictive distribution for the inflation level. We calculate the point forecast $\hat{\pi}_{t+h\mid t}$ by summing the predicted changes from step $1$ to $h$ and adding this cumulative change to the last observed inflation level at time $t$. To quantify the uncertainty associated with this cumulative forecast, we compute the cumulative forecast variance. Under the assumption that forecast errors at each step are independent, the total variance of the $h$-step-ahead forecast, $\hat{\sigma}_h^2$, equals the sum of the individual variances of the predicted changes from step $1$ through $h$. We model the predictive distribution for inflation as
\[
Y_{t+h} \mid \mathcal{F}_t \sim 
\mathcal{N}(\hat{\pi}_{t+h\mid t},\, \hat{\sigma}_h^2).
\]
This specification yields a full probabilistic forecast for inflation.

\subsection{Distributional random forest}
\label{sec:drf}
We use the DRF of \citet{cevid_distributional_2022} as a modern, non-parametric method to model the full conditional distribution of inflation given macroeconomic predictors. The DRF extends the Random Forest (RF) framework of \citet{breiman_random_2001} from conditional mean estimation to conditional distribution estimation. Instead of producing a single point prediction, it constructs an estimator for the entire conditional CDF $F_{Y\mid\mathbf{X}}(y \mid \mathbf{x})$ of $Y$ given $\mathbf{X} = \mathbf{x}$. We therefore consider the DRF particularly suitable for probabilistic forecasting, which requires the full predictive distribution. We implement the DRF in \texttt{R} using the \texttt{drf} package (\url{https://github.com/lorismichel/drf}).

\subsubsection{Forest construction and distributional splitting criterion}
A standard RF builds an ensemble of regression trees to approximate the conditional mean function $\mathbb{E}[Y \mid \mathbf{X}=\mathbf{x}]$. Each tree recursively partitions the feature space and assigns to every terminal leaf a prediction equal to the average response of the training observations in this leaf. The RF prediction for a query point $\mathbf{x}$ is the average of these tree-specific leaf means. We can interpret this aggregation as a local weighting scheme. For any $\mathbf{x}$, the forest induces weights $w_{\mathbf{x}}(i)$ on the training observations $\mathbf{X}_i$, where $w_{\mathbf{x}}(i) > 0$ only if $\mathbf{X}_i$ frequently falls into the same leaves as $\mathbf{x}$. The standard RF estimate of the conditional mean is therefore
\[
\widehat{\mathbb{E}}[Y \mid \mathbf{X}=\mathbf{x}] 
= \sum_{i=1}^n w_{\mathbf{x}}(i) Y_i,
\qquad 
w_{\mathbf{x}}(i) \ge 0,\quad \sum_{i=1}^n w_{\mathbf{x}}(i) = 1.
\]
This interpretation highlights the RF as an adaptive nearest-neighbor method. The weights depend on the local density of the training data and the ensemble structure, and they concentrate mass on observations that are similar to $\mathbf{x}$ in the tree partitions. The DRF extends this framework by using the same forest-induced weights to estimate the entire conditional distribution of $Y$, not just its mean.

Let $\{(Y_i, \mathbf{X}_i)\}_{i=1}^n$ denote the training sample, where $Y_i \in \mathbb{R}$ is the response and $\mathbf{X}_i \in \mathbb{R}^p$ is the predictor vector. Each tree in the DRF recursively partitions the feature space into axis-aligned regions by splitting on one feature and a corresponding threshold. Unlike the standard RF, which chooses splits that maximize differences in conditional means between child nodes, the DRF seeks splits that maximize differences between the entire conditional distributions of $Y$ in the nodes.

Consider a candidate split of the data in one node into two child nodes. The DRF chooses the split that maximizes a measure of distributional dissimilarity between the responses in the two child nodes. The default criterion is the squared maximum mean discrepancy (MMD) \citep{gretton_kernel_2012}, defined for samples $U = \{u_1,\dots,u_{n_U}\}$ and $V = \{v_1,\dots,v_{n_V}\}$ as
\[
\mathrm{MMD}^2(U,V;\psi)
= \frac{1}{n_U^2}\sum_{i,j=1}^{n_U} \psi(u_i,u_j)
+ \frac{1}{n_V^2}\sum_{i,j=1}^{n_V} \psi(v_i,v_j)
- \frac{2}{n_U n_V}\sum_{i=1}^{n_U}\sum_{j=1}^{n_V} \psi(u_i,v_j),
\]
where $\psi(\cdot,\cdot)$ is a characteristic kernel, typically a Gaussian kernel. The MMD compares the empirical kernel mean embeddings of the two distributions. Maximizing $\mathrm{MMD}^2$ at each split encourages child nodes that are internally homogeneous while being as distinct as possible from each other in terms of location, scale, and shape.

\subsubsection{Forest-induced weights and conditional distribution estimation}
After growing $N$ trees, the DRF defines for any query point $\mathbf{x}$ a set of adaptive nearest-neighbor weights based on the ensemble structure. Let $L_b(\mathbf{x})$ denote the set of training indices that are in the same leaf as $\mathbf{x}$ in tree $b$. The forest weight assigned to observation $i$ is
\[
w_{\mathbf{x}}(i)
= \frac{1}{N} \sum_{b=1}^N \frac{\mathbf{1}_{\{ i \in L_b(\mathbf{x}) \}}}{|L_b(\mathbf{x})|},
\qquad
w_{\mathbf{x}}(i) \ge 0, \quad \sum_{i=1}^n w_{\mathbf{x}}(i) = 1.
\]
The weights depend on the similarity between $\mathbf{x}$ and each training point as determined by the tree ensemble. Observations that often share leaves with $\mathbf{x}$ receive higher weights. Their responses $Y_i$ are therefore more informative for predicting the distribution of $Y$ at $\mathbf{x}$. Using these weights, we define the DRF estimator of the conditional CDF as a weighted empirical distribution:
\[
\widehat{F}_{Y\mid\mathbf{X}}(y \mid \mathbf{x}) 
= \sum_{i=1}^n w_{\mathbf{x}}(i) \, \mathbf{1}_{\{ Y_i \le y \}}.
\]
Under regularity conditions, this weighted empirical distribution is a consistent non-parametric estimate of the conditional distribution $F_{Y\mid\mathbf{X}}(y \mid \mathbf{x})$ \citep{cevid_distributional_2022}. Once we have estimated the weights, we can compute any functional of the conditional distribution without retraining, like conditional quantiles and moments:
\[
\widehat{Q}_{Y\mid\mathbf{X}}(\tau \mid \mathbf{x}) 
= \inf\{ y : \widehat{F}_{Y\mid\mathbf{X}}(y \mid \mathbf{x}) \ge \tau \},
\qquad
\widehat{\mathbb{E}}[Y \mid \mathbf{X} = \mathbf{x}] 
= \sum_{i=1}^n w_{\mathbf{x}}(i) Y_i.
\]
The DRF therefore unifies conditional mean and quantile estimation within a single framework and recovers the entire predictive distribution without re-estimation for each functional.

\subsubsection{Forecasting process}
We use the DRF to generate sequential probabilistic inflation forecasts. The forecasting procedure follows the recursive design used for the other models. We first train the DRF on an initial sample comprising the first 40\% of the data. At each monthly forecast origin~$t$, we re-estimate the model on the expanding information set $\mathcal{F}_t$. We then learn the conditional law of future inflation $Y_{t+h}$ for horizons $h \in \{1,3,6,12\}$. Specifically, the DRF estimates the conditional distribution:
\[
F_{t,h}(y) = \mathbb{P}(Y_{t+h} \le y \mid \mathcal{F}_t).
\]
The predictor vector $\mathbf{X}_t$ combines several sources of information (see Appendix~\ref{app:data}). For each macroeconomic variable, we include its current value, its first lag, and its first difference to capture short-term persistence and recent momentum \citep{goulet_coulombe_macroeconomic_2021}. The inflation series itself enters with its current value and eleven autoregressive lags, $\pi_t, \pi_{t-1}, \dots, \pi_{t-11}$, to account for its own dynamics. We use the standard hyperparameters for the DRF, as recommended by the authors. We compute the predictive distribution for $Y_{t+h}$ from the forest weights $\{w_t(i)\}_{i \le t}$. The resulting non-parametric predictive law is
\[
Y_{t+h} \mid \mathcal{F}_t \sim \hat{F}_{t,h}, \qquad
\hat{F}_{t,h}(y) = \sum_{i \le t} w_t(i) \mathbf{1}_{\{\pi_{i+h} \le y\}}.
\]
This distribution provides a probabilistic forecast for inflation at horizon~$h$ and captures both model-based and sampling uncertainty. Repeating this procedure over time yields a sequence $(\hat{F}_{t,h})_t$ of predictive distributions that we evaluate using the methods introduced in Section~\ref{sec:evaluation_theory}.

\end{document}